\newif\ifTwoColumn
\newif\ifTechReport

\TechReporttrue    

\ifTwoColumn
	\documentclass[journal]{IEEEtran}
	\else
		\ifTechReport
		\documentclass[journal,12pt,onecolumn,draftcls]{IEEEtran}
	\else
		\documentclass[journal,11pt,onecolumn,draftclsnofoot]{IEEEtran}
	\fi
\fi

\usepackage[utf8]{inputenc} 
\usepackage[T1]{fontenc}    
\usepackage[hyphens]{url}
\usepackage{booktabs}       
\usepackage{amsfonts,amsmath,amsthm,amssymb,mathrsfs}       
\usepackage{bbm}
\usepackage{cite}
\usepackage{mathtools,xparse}
\usepackage[makeroom]{cancel}
\usepackage[acronym]{glossaries}
\usepackage{glossaries-prefix}
\usepackage{color}
\usepackage{balance}
\usepackage{enumerate}
\usepackage{algpseudocode}
\usepackage{subcaption}
\usepackage{nicefrac}       
\usepackage{microtype}      
\usepackage{xcolor}         
\usepackage{graphicx}
\usepackage{multirow}
\usepackage{booktabs,tabularx}
\usepackage{cases}
\usepackage[ruled,vlined]{algorithm2e}

\usepackage[colorinlistoftodos]{todonotes}

\graphicspath{{figures/}}

\renewcommand{\qedsymbol}{$\blacksquare$}

\DeclareMathSizes{10}{10}{6}{4}

\newcommand{\continuanceref}{}

\newcommand{\R}{\mathbb{R}}

\newcommand{\N}{\mathbb{N}}

\newcommand{\bbS}{\mathbb{S}}
\newcommand{\mc}[1]{\mathcal{#1}}

\newcommand{\eqdef}{\coloneqq}
\newcommand{\reqdef}{\eqqcolon}

\newcommand{\diag}{\mathrm{diag}}

\newcommand{\col}{\mathrm{col}}

\newcommand{\set}[2]{\left\{ #1\ \left| \ #2 \right. \right\}}
\newcommand{\inner}[2]{\langle #1, #2 \rangle}

\newcommand{\bs}[1]{\boldsymbol{#1}}
\newcommand{\bsone}{\boldsymbol{1}}

\newtheorem{theorem}{Theorem}[section]
\newtheorem{definition}[theorem]{Definition}
\newtheorem{proposition}[theorem]{Proposition}
\newtheorem{lemma}[theorem]{Lemma}

\newtheorem{remark}[theorem]{Remark}
\newtheorem{standing}[theorem]{Standing Assumption}

\newcommand{\blue}[1]{\textcolor{blue}{#1}}

\newcommand{\normaltext}[1]{\textnormal{#1}}

\makeglossaries
\newacronym[
    prefixfirst={a\ },
    prefix={an\ }
]{LP}{LP}{linear program}
\newacronym[
    prefixfirst={a\ },
    prefix={an\ }
]{LTI}{LTI}{linear time-invariant}
\newacronym{QP}{QP}{quadratic program}

\newglossaryentry{LMI}
{
	name={LMI},
	description={linear matrix inequality},
	first={\glsentrydesc{LMI} (\glsentrytext{LMI})},
	plural={LMIs},
	descriptionplural={linear matrix inequalities},
	firstplural={\glsentrydescplural{LMI} (\glsentryplural{LMI})}
}

\newglossaryentry{BMI}
{
	name={BMI},
	description={bilinear matrix inequality},
	first={\glsentrydesc{BMI} (\glsentrytext{BMI})},
	plural={BMIs},
	descriptionplural={bilinear matrix inequalities},
	firstplural={\glsentrydescplural{BMI} (\glsentryplural{BMI})}
}

\newacronym{SDP}{SDP}{semidefinite program}
\newacronym{iid}{i.i.d.}{independent and identically distributed}
\newacronym{wrt}{w.r.t.}{with respect to}
\newacronym{KKT}{KKT}{Karush-Kuhn-Tucker}

\newacronym{GNE}{GNE}{generalized Nash equilibrium}
\newacronym{v-GNE}{v-GNE}{variational generalized Nash equilibrium}
\newacronym{f-GNE}{f-GNE}{feedback-optimized generalized Nash equilibrium}
\newacronym{e-GNEP}{e-GNEP}{externally-guided generalized Nash equilibrium problem}
\newacronym{e-GNE}{e-GNE}{externally-guided generalized Nash equilibrium}
\newacronym{GNEP}{GNEP}{generalized Nash equilibrium problem}
\newacronym{f-GNEP}{f-GNEP}{feedback-optimized generalized Nash equilibrium problem}
\newacronym{FB}{FB}{forward-backward}
\newacronym{pFB}{pFB}{preconditioned forward-backward}
\newacronym{IQC}{IQC}{integral quadratic constraint}
\newacronym{VI}{VI}{variational inequality}

\newacronym{DSO}{DSO}{distribution system operator}
\newacronym{LFM}{LFM}{local flexibility market}
\newacronym{DER}{DER}{distributed energy resource}
\newacronym{EV}{EV}{electric vehicle}
\newacronym{ES}{ES}{energy storage}
\newacronym{AS}{AS}{ancillary service}

\begin{document}


\title{Incentives and co-evolution: Steering linear dynamical systems with noncooperative agents}
\author{Filippo Fabiani and Andrea Simonetto
	\thanks{F. Fabiani is with the IMT School for Advanced Studies Lucca, Piazza San Francesco 19, 55100 Lucca, Italy ({\tt filippo.fabiani@imtlucca.it}). A. Simonetto is with the Unité de Mathématiques Appliquées, {ENSTA Paris}, Institut Polytechnique de Paris, 91120 Palaiseau, France, {\tt \footnotesize (andrea.simonetto@ensta-paris.fr)}.}}


\maketitle

\begin{abstract}
	Modern socio-technical systems
	typically consist of many interconnected users and competing service providers, where
	notions like market equilibrium are tightly connected to the ``evolution'' of the network of users. In this paper, we model the users' dynamics as a linear dynamical system, and the service providers as agents taking part to a generalized Nash game, whose outcome coincides with the input of the users' dynamics. We thus characterize the notion of co-evolution of the market and the network dynamics and derive dissipativity-based conditions leading to a pertinent notion of equilibrium. 
	We then focus on the control design and adopt the \emph{light-touch} policy to incentivize or penalize the service providers \emph{as little as possible}, while steering the networked system to a desirable outcome.
	We also provide a dimensionality-reduction procedure, which offers network-size independent conditions. Finally, we illustrate our novel notions and algorithms on a simulation setup stemming from digital market regulations for influencers, a topic of growing interest. 	
\end{abstract}


\IEEEpeerreviewmaketitle

\glsresetall

\section{Introduction}\label{sec:intro}
\IEEEPARstart{M}{\lowercase{odern}} cyber-physical and social systems, as smart grids, ride-hailing services, or digital marketplaces, are typically composed of many interconnected users (or customers) and competing service providers (or agents) that mutually influence each other. 
Building upon this tight connection, we are interested in modeling, analyzing, and stabilizing the closed-loop system  between the competing providers who influence (and are influenced by) the users, and the users who ``evolve'' accordingly. Specifically, we model the users' dynamics as governed by a \gls{LTI}  dynamical system, and the service providers as decision-making agents taking part to a generalized Nash game (hereinafter also called \gls{GNEP} with a slight abuse), whose outcome coincides with the input of the users' dynamics. 

%
The study of multi-agent systems involving these type of heterogeneous interactions is receiving growing attention in the last few years. Prominent examples can be found in digital platforms and recommender systems \cite{Rossi2022,Jagadeesan2022,Jiang2019}, where the latter adapt their output to the reactions of the users who are, in turn, affected by the recommended content, and closed-loop machine learning paradigms \cite{ivano2022,D'Amour2020}, which study long-term behaviours of deployed machine learning-based decision systems by accounting for their potential future consequences through notions of fairness, equitability, or other ethical concepts. 
Our work is indeed strongly motivated by digital marketplaces, in particular the problem of regulating the advertising market involving competitive firms and influencers in social networks, briefly introduced next and further elaborated in \S \ref{sec:case_study}.
\begin{figure}
	\centering
	\includegraphics[width=0.4\textwidth]{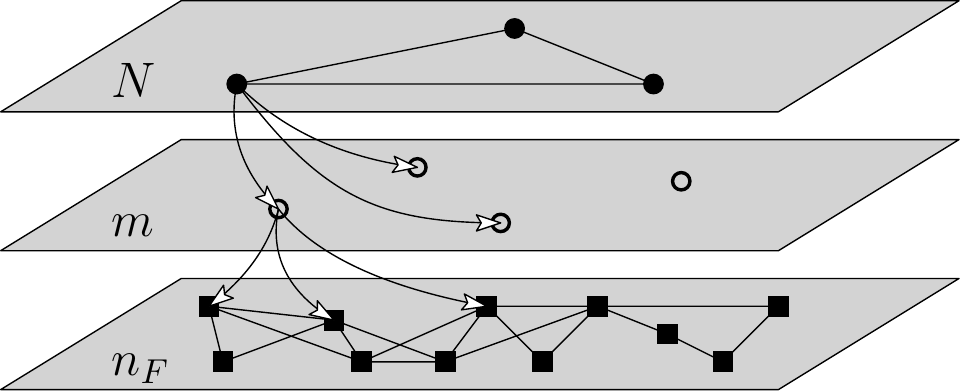}
	\caption{Multi-level interactions for the application considered in this work (see also \S \ref{sec:case_study}): $N$ firms paying $m$ influencers, who in turn influence the dynamics of a population of $n_F$ followers.}
	\label{cool.pic}
\end{figure}
	\subsection{Motivating example: market regulation on social networks}\label{subsec:motivating_examples}
	A leading role in advertising markets is nowadays personified by social influencers who post videos and photos in social networks featuring sponsored contents and advertisements. The market value is estimated at over 16B~{\sc{usd}}, with over 100M influencers of different ``size'', and roughly 20\% of companies investing half of their annual marketing budget to it, and it can not be left unregulated especially for large influencers~\cite{Narassiguin2019, Stats2022, Influencer2022}.
	
	To fix the ideas, we will hence consider $m$ influencers who recommend products to a population of followers. The former are then directly ``exploited'' by $N$ competitive firms, which aims at selling a desirable quantity of products guaranteeing a certain degree of profit, and hence invest their money $y_i$ to pay the influencers to advertise them. Social influencers, on their side, are connected through, e.g., social networks, with a population of $n_F$ followers (i.e., the consumers), and hence can steer the sale of those products throughout the network. The followers' state $x_F$ may indeed represent how much of a certain product people buy -- see Fig.~\ref{cool.pic} for a pictorial representation. 
	
	Mathematically speaking, we assume each company participates in a generalized Nash game, whose outcome is their optimal investment $y_i^*(x_F)$, a function of the current state of the customers $x_F$. Such investment is used to steer (by paying influencers) the consumers to buy their products. On the other hand, we model the consumers' purchasing evolving as an \gls{LTI} system, whose input is the advertisement strength they receive and their dynamics is affected by their peers via social bonds (exemplified by a network in Fig.~\ref{cool.pic}). The state evolution of $x_F$ to $x_F^+$ determines another optimal investment strategy $y_i^*(x_F^+)$ and so on. Customers and companies are then coupled through a closed-loop system and thereby co-evolving. We are interested here in characterizing a pertinent notion of equilibrium of such system, and designing control laws, i.e., taxation and regulation schemes to shape $y_i^*(x_F)$, to steer the underlying dynamics to co-evolve towards a desirable equilibrium, satisfying all the main actors involved in the networked system. Our taxation (or, in some cases, incentives) are directly applied by the government to the influencers revenues, thereby reducing the leverage companies have on them, thus curtailing the influencers' effect on the customers.  
	
	



Besides the motivating example described above and revisited later in \S \ref{sec:case_study}, we stress that the way we will model the overall co-evolutionary problem, perform the resulting analysis and control design is fairly general and covers also different scenarios, as for instance the intrinsic competition and bi-level interactions arising in energy flexibility markets \cite{heinrich2021local,evangelopoulos2022heterogeneous}.
As such, we are interested in market incentives and regulations, whereby firms compete each other to influence users to buy certain products, and the users evolve following a suitably defined dynamics incorporating network effects and external inputs.

%
%
%
\subsection{Related work and summary of contribution}

Our work investigates the joint evolution of a set of agents taking part to a \gls{GNEP}, whose outcome influence (and it is influenced by) a \gls{LTI} dynamics underlying interconnected agents. Unlike available results in algorithmic game theory \cite{paccagnan2016distributed,pavel2019distributed,yi2019operator,belgioioso2022distributed}, however, we do not propose any \gls{GNE} seeking scheme, since we are interested in the analysis and control of the interconnected system as a whole, thereby aiming at reaching a \textit{co-evolutionary equilibrium} (\S \ref{sec:problem_description}). In this sense, our work also differs from recent papers proposing MPC-inspired game-theoretic schemes, such as \cite{muller2017economic,fele2018framework,hall2022receding}. In our framework, in fact, noncooperative agents and \gls{LTI} system are treated as separate, yet mutually coupled, entities, which shall be driven towards some operational condition that is desirable for the overall networked system.

The technical results developed in the paper (\S \ref{sec:synthesis_analysis}, \ref{sec:controller_design}) borrow tools from standard dissipativity theory and, specifically, from \cite{megretski1997system,lessard2016analysis}. Similar techniques have also recently been employed in a purely game-theoretic context, for example to establish asymptotic stability of the set of Nash equilibria for deterministic population games, combining  payoff and evolutionary dynamics models  \cite{Arcak2021}, or to analyze the convergence properties of (typically, continuous-time) \gls{GNE} seeking procedures \cite{Pavel2022}.

Bearing in mind the case study involving firms, influencers and potential costumers presented in \S \ref{sec:case_study}, we note that the proposed control methodology (\S \ref{sec:controller_design}) can be thought of as an incentive/charging design paradigm, especially the part based on the light-touch principle. Suitable examples can be found, for instance, in \cite{fabiani2021personalized,fabiani2022learning,Yan2022,Yan2023hierarchical}. While in \cite{fabiani2021personalized,fabiani2022learning} the design of personalized incentives enabled for the distributed computation of a \gls{GNE}, \cite{Yan2022} proposed a Pareto-based incentive mechanism under sustainable budget constraint to improve the social welfare of the agents taking part to a game, where a central coordinator redistributes collected taxes among the population in order to remodel agents’ dynamical decision-making. A social welfare improvement was also considered in \cite{Yan2023hierarchical}, where intra-group incentives were designed to stabilize dynamical agents to the group Nash equilibrium in a hierarchical framework.

Closer in the spirit to the problem considered in this paper are those works concerning recommender systems \cite{Rossi2022,Jagadeesan2022,Jiang2019}, and those falling within the social network and dynamic opinion formation literature, as \cite{Friedkin2015,fontan2017multiequilibria,Proskurnikov2017,Proskurnikov2018}, also possibly accompanied by some form of control, influence, or nudging \cite{Acemoglu2011,Perra2019}. Compared to the aforementioned works, however, a crucial difference is represented by the proposed modeling paradigm, and subsequent analysis and control synthesis, which includes the notion of agents competing to influence some \gls{LTI} dynamics, and an external entity regulating the overall market.

In summary, our paper makes the following contributions:
\begin{itemize}
	\item We model the networked system made by a set of selfish agents taking part to a \gls{GNEP} whose outcome affects (and is affected by) the evolution of some \gls{LTI} system, and we introduce a novel notion of equilibrium for it;
	\item Focusing on the stability analysis of the networked system, we establish easy-to-check sufficient conditions based on \gls{LMI} guaranteeing asymptotic convergence to a co-evolutionary equilibrium;
	\item We develop \glspl{BMI} for the control synthesis, which can be solved efficiently via a proposed bisection-like method if one relies on the newly investigated light-touch principle for the control design;
	\item To alleviate the computational burden for \gls{LTI} systems with many states, we provide a dimension-reduction procedure offering network-size independent conditions. Even if such a procedure is developed on our case study with light-touch policy, it is general and can be applied to design a variety of controllers meeting the required conditions;
	\item As a case study, we develop a novel model involving the digital market regulation for influencers paid by companies to advertise their products in order to attract customers.
\end{itemize}

The proofs of theoretical results are all deferred to Appendix.


\subsection*{Notation}
$\N$, $\R$, and $\R_{\geq 0}$ denote the set of natural, real, and nonnegative real numbers, respectively. $\N_0 \eqdef \N \cup \{0\}$. $\bbS^{n}$ is the space of $n \times n$ symmetric matrices and $\bbS_{\succ 0}^{n}$ ($\bbS_{\succcurlyeq 0}^{n}$) is the cone of positive (semi)definite matrices.
The transpose of a matrix $A \in \R^{n \times n}$ is $A^\top$, $\Lambda(A)$ the set of its eigenvalues $\{\lambda_1, \ldots, \lambda_n\}$ with $\lambda_\textrm{max} \eqdef \textrm{max}_{i = 1, \ldots, n} \, \{\lambda_i\}$, and $[A]_{ij}$ its $(i,j)$-th entry. 
$A \otimes B$ is the Kronecker product between matrices $A$ and $B$. 
$A \succ 0$ ($\succcurlyeq 0$) stands for a positive (semi)definite matrix. 
Given a vector $v \in \R^n$ and a matrix $A \in \bbS^n$, we denote with $\|v\|$ the standard Euclidean norm, while with $\|\cdot\|_A$ the $A$--induced norm such that $\|v\|_A \coloneqq \sqrt{v^\top A v} = \sqrt{\inner{Av}{v}}$, where $\inner{\cdot}{\cdot} : \mathbb{R}^n \times \mathbb{R}^n \rightarrow \mathbb{R}$ stands for the standard inner product. $\mc{B}_\theta \eqdef \set{x \in \R^n}{\|x\| \leq \theta}$.
$I_{n}$, $\bsone_n$, $\bs{0}_n$ denote the $n \times n$ identity matrix, the vector of all $1$ and $0$, respectively (we omit the dimension $n$ whenever clear from the context). The uniform distribution on the closed interval $[a,b]$ is denoted by $\mathcal U(a,b)$. 
The operator $\col(\cdot)$ (resp., $\diag(\cdot)$) stacks its arguments in column vectors or matrices (block-diagonal matrix) of compatible dimensions. To indicate the state evolution of discrete-time \gls{LTI} systems, we use $x_{k+1}$, $k \in \N_0$, as opposed to $x^+$, to make the time dependence explicit whenever necessary.


For ease of visualization, we highlight in \blue{blue font} the decision variables in the matrix inequalities developed throughout.

\subsubsection{Operator-theoretic definitions (\hspace{-.03mm}\textup{\cite{bauschke2011convex}})}
Given a nonempty and convex set $\mc{X} \subseteq \R^n$, $T : \mc X  \rightrightarrows \mathbb{R}^n$ is monotone if $\inner{T(x) - T(y)}{x - y} \geq 0$ for all $x, y \in \mc{X}$, and it is $\mu$-strongly monotone, $\mu > 0$, if $\inner{T(x) - T(y)}{x - y} \geq \mu \|x - y\|^2$, for all $x, y \in \mc{X}$.

\subsubsection{Variational inequality (\hspace{-.03mm}\textup{\cite{facchinei2007finite}})} 
A \gls{VI} is defined by a feasible set $\mc{X} \subseteq \R^n$, and a mapping $F : \mc{X} \to \R^n$. We denote by VI$(\mc{X}, F)$ the problem of finding some vector $x^* \in \mc{X}$ such that $(y - x^*)^\top F(x^*) \geq 0, \, \text{ for all } y \in \mc{X}$. Such an $x^*$ is therefore called a \textit{solution} to VI$(\mc{X}, F)$, and the associated set of solutions is denoted as $\mc{S} \subseteq \mc{X}$.

\subsubsection{Graph theory (\hspace{-.03mm}\textup{\cite{mesbahi2010graph}})}
Let $\mathcal{G} = (\mc V, \mc E)$ be an undirected graph connecting a set of vertices $\mc V = \{1,\ldots,V\}$ through a set of edges  $\mc E \subseteq \mathcal{V} \times \mathcal{V}$, with $|\mathcal{E}| = E$ and $(i, j) \in \mc E$ only if there is a link connecting nodes $i$ and $j$. The set of neighbours of node $i$ is defined as $\mathcal{N}_i = \{j \in \mc V \mid (i,j) \in \mathcal{E}\}$. The graph $\mathcal{G}$ is connected if there exists a sequence of distinct nodes such that any two subsequent nodes form an edge between any two vertices of $\mathcal{G}$. 
To define the incidence matrix $D \in \mathcal{R}^{E \times V}$ associated to $\mathcal{G}$, we label the edges $e_l \in \mc E$ for $l = \{1,\ldots, E\}$ considering an arbitrary orientation, yielding $[D]_{li} = -1$ if $i$ is the output vertex of $e_l$, $[D]_{li} = 1$ if $i$ is the input vertex of $e_l$, $[D]_{li} = 0$ otherwise. By construction, $D \bs{1}_V =
\bs{0}_E$, and if $\mathcal{G}$ is connected, $D x = \bs{0}_E$ if and only if $x \in \{\alpha \bs{1}_V \mid \alpha \in \R\}$.
We denote by $L \in \mathbb{R}^{V \times V}$ the Laplacian matrix of the graph $\mathcal{G}$, with $[L]_{ij} = |\mathcal{N}_i|$ if $i=j$, $[L]_{ij} = -1$ if $(i,j) \in \mathcal{E}$, $[L]_{ij} = 0$ otherwise. Additionally, it holds that $L = D^\top D$.

\glsresetall

\section{Problem description and preliminaries}\label{sec:problem_description}
We start by introducing the mathematical model considered and related technical discussion, which will be instrumental for its analysis and subsequent controller(s) synthesis.

\subsection{Mathematical formulation}\label{subsec:model}
We investigate the dynamical evolution and closed-loop properties of the system obtained by interconnecting a population of agents taking part to a \gls{GNEP} whose outcome is affected by the state variables of a certain discrete-time \gls{LTI} system.

Specifically, we consider a noncooperative game involving $N$ agents, indexed by the set $\mc I \eqdef \{1, \ldots, N\}$, each one taking (locally constrained) decisions $y_i \in \mc{Y}_i \subseteq \R^{p_i}$ to minimize some local cost function  while sharing, and therefore competing for, limited resources with the other agents.
Unlike traditional \glspl{GNEP}, however, we assume that both the cost function of each agent and the coupling constraints depend not only on the decisions of the other agents $\bs{y}_{-i} \eqdef \col((y_j)_{j \in \mc I \setminus \{i\}}) \in \R^{p -p_i}$, $p \eqdef \sum_{i \in \mc I} p_i$, but also on some external variable $x \in \R^n$ that can be likewise influenced by the collective decision vector $\bs{y} \eqdef \col((y_i)_{i \in \mc I}) = (y_i, \bs{y}_{-i}) \in \R^p$ through some control input $u \in \R^m$. If we hence let $x$ being governed by a \gls{LTI} dynamics through some pair of system matrices $A \in \R^{n \times n}$ and $B \in \R^{n \times m}$, we now describe the \gls{GNEP} with external influence $\Gamma \eqdef (\mc I, (J_i)_{i \in \mc I}, (\mc{Y}_i)_{i \in \mc I}, (A,B))$ at hand by means of the following collection of optimization problems:
\begin{equation}\label{eq:single_prob}
	\forall i \in \mc I : \left\{
	\begin{aligned}
		& \underset{y_i \in \mc{Y}_i}{\textrm{min}} && J_i(y_i, \bs{y}_{-i}, x)\\
		& \textrm{ s.t. } && (y_i, \bs{y}_{-i}) \in \Omega(x),
	\end{aligned}
	\right.
\end{equation}
where $J_i : \R^p \times \R^n \to \R$ denotes the local cost function of each agent, $\Omega : \R^n \to 2^{\R^l}$ the set of state-dependent constraints coupling the decisions of the $N$ agents, while the variable $x$ is constrained to some $\mc{X} \subseteq \R^n$ and evolves as follows:
\begin{equation}\label{eq:linear_dynamics}
	x^+ = A x + B u .\\
\end{equation}
Given that the state variable $x$ appears both in the cost function and constraints in \eqref{eq:single_prob}, we note that each local decision (and hence also the collective one $\bs y$) is actually a function of $x$ itself, i.e., $y_i = y_i(x)$ -- we will make this dependency explicit or omit it according to the context.
In the remainder we assume the agents are competing with each other for controlling the dynamical system \eqref{eq:linear_dynamics}, and in particular the state $x$.  Specifically, each agent taking part to the \gls{GNEP} has a desired set point $\bar{x}_i \in \mc{X}$ for \eqref{eq:linear_dynamics} and has available some ``resources'', which without restriction may coincide with $y_i$ itself, to influence \eqref{eq:linear_dynamics} through $u$. This notion will be formalized later in \S \ref{sec:synthesis_analysis}.
%

\sloppy After introducing sets $\mc{Y} \eqdef \prod_{i \in \mc I} \mc{Y}_i \subseteq \R^p$, $\mc{Y}_i(\bs{y}_{-i},x) \eqdef \set{z \in \mc{Y}_i}{(z, \bs{y}_{-i}) \in \Omega(x)}$, in the considered framework we are then interested in the following notion of equilibrium:
\begin{definition}\textup{(Co-evolutionary equilibrium)}\label{def:e_GNEP}
	A pair $(x^*,\bs{y}^*(x^*)) \in \mc{X} \times \R^p$ is a \emph{co-evolutionary equilibrium} for the \normaltext{\gls{GNEP}} $\Gamma$ in \eqref{eq:single_prob} and discrete-time \normaltext{\gls{LTI}} system in \eqref{eq:linear_dynamics} if \normaltext{i)} $B u^* = (I-A) x^*$ for some $u^* \in \R^m$, and \normaltext{ii)} we have
	\begin{equation}\label{eq:standard_GNE}
		J_i(y_i^*, \bs{y}_{-i}^*, x^*) \le \underset{\xi_i \in \mc{Y}_i(\bs{y}^*_{-i}, x^*)}{\mathrm{inf}} \ J_i(\xi_i, \bs{y}_{-i}^*, x^*) ,
	\end{equation}
	for all $i \in \mc I$. \hfill$\square$
\end{definition}

\begin{figure}[t!]
	\centering
	\ifTwoColumn
\else
		\includegraphics[width=.6\columnwidth]{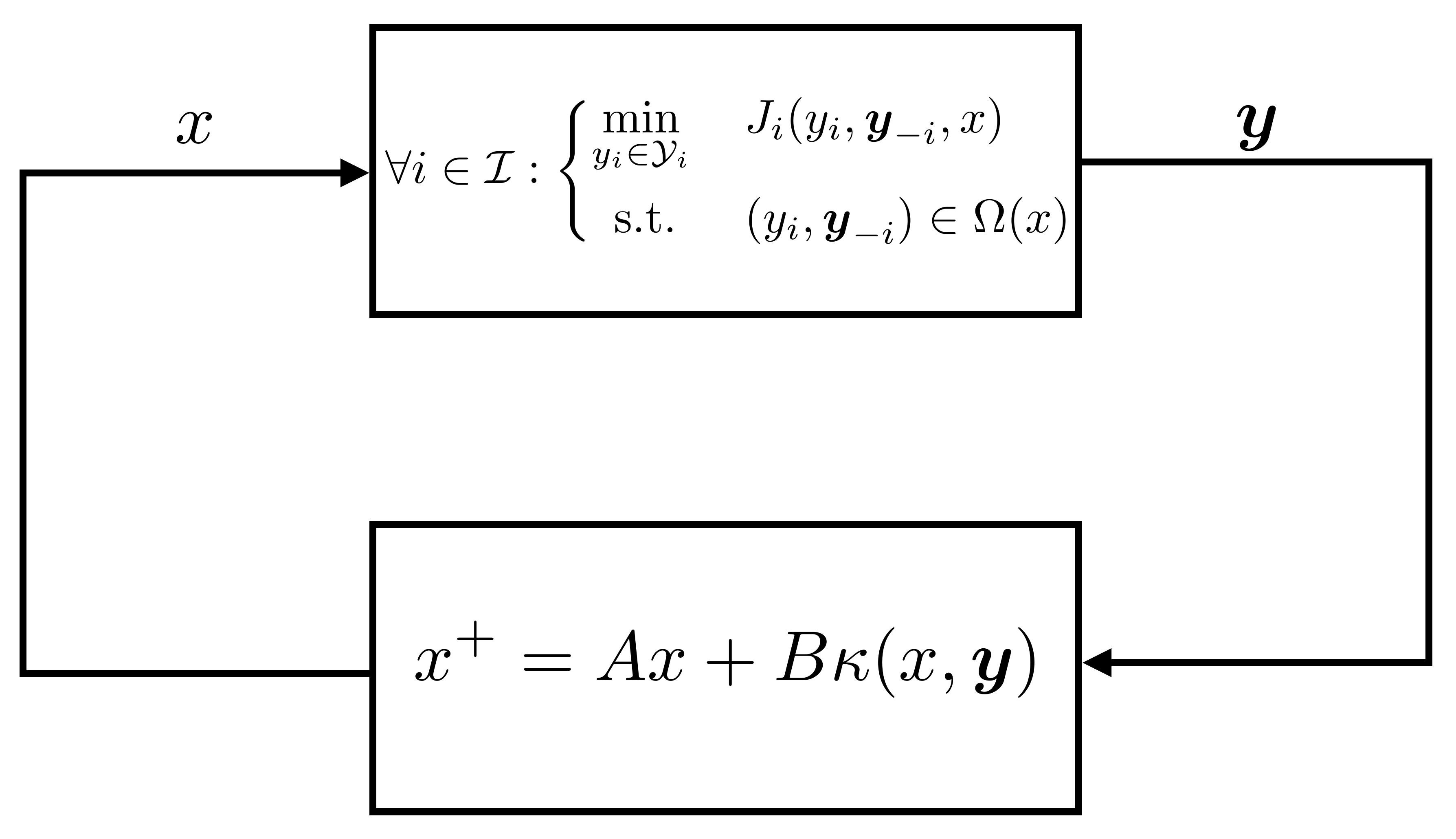}
	\fi	
	\caption{Networked system consisting of a population of agents involved in a \gls{GNEP}, whose outcome is affected by the evolution of a discrete-time \gls{LTI} system.}
	\label{fig:control_loop}
\end{figure}

Our goal is hence to find a suitable control law $\kappa : \R^n \times \R^p \to \R^m$, possibly dependent (either directly or implicitly) on  both the state $x$ of the system \eqref{eq:linear_dynamics} and the collective profile $\bs{y}$, so that $u = \kappa(x, \bs y(x))$ asymptotically drives the closed-loop networked system to a co-evolutionary equilibrium, while satisfying both state and state-dependent constraints $\Omega(\cdot)$. See Fig.~\ref{fig:control_loop} for a pictorial representation of the whole system.

With this regard, the first condition stated in Definition~\ref{def:e_GNEP} shall be satisfied with some $u^* = \kappa(x^*, \bs{y}^*(x^*))$, thus turning into $B \kappa(x^*, \bs{y}^*(x^*)) = (I-A) x^*$, namely the pair $(\kappa(x^*, \bs{y}^*(x^*)), x^*)$ identifies a valid steady-state solution for the dynamics in \eqref{eq:linear_dynamics}. Specifically, this requires one to find a feasible collective vector of strategies $\bs{y}^*(x^*)$ that leads the system in \eqref{eq:linear_dynamics} to an equilibrium $x^* \in \mc{X}$ and fits the standard notion of \gls{GNE} when $x = x^*$ in \eqref{eq:single_prob}. 
Meeting both conditions simultaneously is crucial. In fact, given a certain $\bs{y}^*(\bar x) \in \Omega(\bar x)$ satisfying \eqref{eq:standard_GNE} for some $\bar{x} \in \mc{X}$, in case this latter does not allow to make $B \kappa(\bar{x}, \bs{y}^*(\bar x)) = (I-A) \bar{x}$ true, then the \gls{LTI} system \eqref{eq:linear_dynamics} evolves to a different point, thus possibly invalidating the current \gls{GNE} $\bs{y}^*(\bar x)$. If there exists, instead, a feasible collective strategy $\bar{\bs{y}}(x^*)$ leading to some $x^* \in \mc{X}$ so that $B \kappa(x^*, \bar{\bs{y}}(x^*)) = (I-A) x^*$ is verified though \eqref{eq:standard_GNE} is not, then some of the agents can improve their cost by deviating from $\bar{\bs{y}}(x^*)$, which hence results in an inefficient strategy profile. 

\subsection{Technical preliminaries}\label{sec:technical}
First, we make some assumptions that will hold throughout:
\begin{standing}\label{standing:basic}
	The following conditions hold true:
	\begin{enumerate}
		\item[\normaltext{(i)}] For each $i \in \mc I$, $J_i(\cdot, \bs{y}_{-i},x)$ is a $\mc C^1$, convex function, for fixed $\bs{y}_{-i} \in \mc{Y}_{-i}$ and $x \in \mc{X}$;
		\item[\normaltext{(ii)}] For each $i \in \mc I$, $\mc{Y}_i$ is a nonempty, compact, and convex set. For every $x \in \mc{X}$, the set $\Omega(x)$ is nonempty and $\Omega(x) \cap \mc Y$ satisfies the Slater’s constraint qualification.
		\hfill$\square$
	\end{enumerate}
\end{standing}

The conditions stated in Standing Assumption~\ref{standing:basic} typically guarantee the existence of at least a \gls{GNE} for the \gls{GNEP} in \eqref{eq:single_prob} with fixed $x$ -- see, e.g., \cite[Ch.~12]{palomar2010convex}.
Moreover, by referring to \eqref{eq:single_prob} for a fixed state $x$, agents typically compute a so-called \gls{v-GNE}. Remarkably, such problem is equivalent to solve $\textrm{VI}(\Omega(x) \cap \mc Y, F(\cdot, x))$ \cite{facchinei2007generalized} where, in view of Standing Assumption~\ref{standing:basic}.(i), $F : \R^p \times \R^n \to \R^p$  is a continuously differentiable single-valued mapping defined as $F(\bs{y},x) \eqdef \col((\nabla_{y_i} J_i(y_i, \bs{y}_{-i}, x))_{i \in \mc I})$. In this way, since $\Omega(x) \cap \mc Y$ is assumed nonempty for any $x \in \mc X$, the set of \gls{v-GNE} is nonempty as well and coincides with the set-valued mapping $\mc{S} : \R^n \rightrightarrows \R^p$ defined as
\ifTwoColumn
\else
	\begin{equation}\label{eq:vGNE}
		\mc{S}(x) \eqdef \set{\bs{y} \in \Omega(x)\cap\mc Y}{(\bs{z} - \bs{y})^\top F(\bs{y},x) \ge 0, \textrm{ for all } \bs{z} \in \Omega(x)\cap\mc Y} .
	\end{equation}
\fi

We next assume additional properties on the mapping $F(\cdot,x)$ that will allow us to claim uniqueness of the \gls{v-GNE} for any fixed $x$, i.e., $\mc{S}(x)$ turns out to be a singleton \cite[Ch.~12]{palomar2010convex}:
\begin{standing}\label{standing:strong_monotonicity}
	The pseudo-gradient mapping $F : \R^p \times \R^n \to \R^p$  satisfies the following conditions:
	\begin{enumerate}
		\item[\normaltext{(i)}] For any fixed $x \in \mc{X}$, $F(\cdot, x)$ is $\eta$-strongly monotone and $\ell$-Lipschitz continuous, for $\eta$, $\ell > 0$;
		\item[\normaltext{(ii)}] For any fixed $\bs{y} \in \mc{Y}$, $F(\bs{y}, \cdot)$ is differentiable, and $\textnormal{\textrm{sup}}_{x \in \mc{X}, \bs{y}\in \Upsilon } \, \|\nabla_{x} F(\bs{y}, x)\| \leq \theta$, for $\theta > 0$.
		\hfill$\square$
	\end{enumerate}
\end{standing}

Note that the strong monotonicity assumption is quite standard in algorithmic game theory \cite{paccagnan2016distributed,yi2019operator,belgioioso2018projected}. In view of the postulated conditions, our problem therefore reduces to finding a feedback law $\kappa(x, \bs{y})$ that allows us to meet the following set of steady-state and equilibrium conditions:
$$
\left\{
\begin{aligned}
	& B \kappa(x^*, \bs{y}^*(x^*)) = (I-A) x^* ,\\
	& x^*  \in \mc{X} ,\\
	& \bs{y}^*(x^*) \in \mc{S}(x^*) .
\end{aligned}
\right.
$$
We derive next a technical result characterizing $\mc S(\cdot)$ and $\bs{y}(\cdot)$:
\begin{lemma}\label{lemma:properties}
	The following statements hold true:
	\begin{enumerate}
		\item[\normaltext{(i)}] For all $x \in \mc{X}$, $\mc{S}(x)$ is a singleton;
		\item[\normaltext{(ii)}] For all $x$, $x' \in \mc{X}$, $\|\bs{y}^*(x) - \bs{y}^*(x')\| \leq \frac{\theta}{\eta} \|x - x'\| $.
		\hfill$\square$
	\end{enumerate}
\end{lemma}

We stress that the nonmonotonicity of $F$ due to the coupling between $\bs y$ and $x$, the current generic structure of the controller $\kappa$, along with the presence of state constraints $\mc X$ acting on the \gls{LTI} dynamics in \eqref{eq:linear_dynamics}, 
complicate the analysis of the networked system, which hence requires tailored tools and control solutions to govern the resulting joint evolution. These are the main topics covered within the next two sections.


\section{Closed-loop analysis of the networked system}\label{sec:synthesis_analysis}
\subsection{Preliminary discussion}
We start our analysis imposing further assumptions on the structure of the control action $\kappa$ and cost functions in \eqref{eq:single_prob}. 
As common in control theory, we thus require the controller $\kappa$ to be linear in the agents' collective strategy, which on the other hand implicitly depends on the state variable $x$, thus resulting into a nonlinear controller for the system in \eqref{eq:linear_dynamics}:
\begin{equation}\label{eq:controller}
	\kappa(x,\bs{y}) = \sum_{i \in \mc I} K_i y_i(x) = K \bs{y}(x) ,
\end{equation}
with suitable gains $K_i \in \R^{m \times p_i}$ to be designed, $K \eqdef [K_1 \ K_2 \, \cdots \, K_N] \in \R^{m \times p}$. 
In addition, we consider each cost function in \eqref{eq:single_prob} to be taken in the following form: 
\begin{equation}\label{eq:costfun_new}
	J_i(y_i, \bs{y}_{-i}; x) \eqdef  \frac{1}{2}\|A x + B K \bs{y} - \bar{x}_i \|^2_{Q_i} + f_i(y_i, \bs{y}_{-i}) ,
\end{equation}
for $Q_i \succ 0$ and $f_i : \R^p \to \R$ chosen so that Standing Assumption~\ref{standing:basic} is met, for all $i \in \mc I$. 
In particular, the presence of each $\bar{x}_i \in \mc{X}$ in the cost, and more generally of the term $\|A x + B K \bs{y} - \bar{x}_i \|^2_{Q_i}$, reflects the willingness of each agent taking part to the \gls{GNEP} to steer the \gls{LTI} system in \eqref{eq:linear_dynamics} to some desired set point, for which it invests available ``resources'' $y_i$, which therefore appear linearly in the control action $\kappa$ in \eqref{eq:controller}.

Thus, the problem we want to solve translates into finding some (possibly constrained) controller gain matrix $K$ such that the coupled generalized Nash game with \gls{LTI} system: 
$$
	\forall i \in \mc I : \left\{
	\begin{aligned}
		& \underset{y_i \in \mc{Y}_i}{\textrm{min}} && \frac{1}{2}\|A x + B K \bs{y} - \bar{x}_i \|^2_{Q_i} + f_i(y_i, \bs{y}_{-i})\\
		&\hspace{.1cm}\textrm{ s.t. } && (y_i, \bs{y}_{-i}) \in \Omega(x),
	\end{aligned}
	\right.
$$

$$
	\hspace{-2cm}x^+ = A x + B K \bs{y} ,
$$
reaches a co-evolutionary equilibrium in the sense of Definition~\ref{def:e_GNEP}. 
In other words, we want to design suitable incentives $K \bs{y}$ to drive the \gls{LTI} system to an equilibrium that is compatible with the selfish agents desires $\bar{x}_i$, while co-evolving with it. 

In the considered setting, i.e., with cost functions as in \eqref{eq:costfun_new}, the pseudo-gradient mapping hence reads as 
\begin{eqnarray*}
&&\!\!\!\!\!\!F(\bs{y},x) = \col((\nabla_{y_i} f_i(y_i, \bs{y}_{-i}) + \\ && \qquad\qquad \qquad K_i^\top B^\top Q_i (Ax + B K \bs y - \bar{x}_i))_{i \in \mc I}) \\
&&\!\!\!\!\!\!\nabla_x F = \diag((K_i)_{i \in \mc I})^\top (B^\top \otimes I_N) \col((Q_i)_{i \in \mc I}) A \in \R^{p \times n}.
\end{eqnarray*}
Furthermore, we note that $\theta = \theta(K) \leq \|\col((Q_i)_{i \in \mc I}) A\|\|(B \otimes I_N) \diag((K_i)_{i \in \mc I})\|$, while the strong monotonicity and Lipschitz constant coefficients $\eta$ and $\ell$ characterizing $F$ also depend on the choice of $f_i(\cdot)$. Thus, given some equilibrium $x^* \in \mc{X}$ for \eqref{eq:linear_dynamics}, in view of Lemma~\ref{lemma:properties}.(ii) for any $x \in \mc{X}$ we have $\|\bs{y}^*(x) - \bs{y}^*(x^*)\| \leq \frac{\theta}{\eta} \|x - x^*\|$, which directly leads to the following dissipative-like condition:
	\begin{equation}\label{eq:IQC}
		\begin{bmatrix}
			\bs{y}^*(x) - \bs{y}^*(x^*) \\ x - x^*
		\end{bmatrix}^\top \begin{bmatrix} -I & \bs 0 \\ \phantom{-}\bs 0 &  (\theta/\eta)^2 I \end{bmatrix} \begin{bmatrix} \bs{y}^*(x) - \bs{y}^*(x^*) \\ x - x^* \end{bmatrix} \geq \bs 0 .
	\end{equation}

\begin{algorithm}[!t]
	\caption{Two-timescale procedure}\label{alg:two_layer}
	\DontPrintSemicolon
	\SetArgSty{}
	\SetKwFor{ForAll}{for all}{do}{end forall}
	\smallskip
	\textbf{Initialization:} $x_0 \in \mc X$\\
	\smallskip
	\textbf{Iteration $(k \in \N_0)$:}
	\vspace{-.1cm}
	$$
	\begin{array}{clr}
		&\bs{y}_k = \mathsf{GNE}(x_k) & \text{(\gls{GNE} computation)}\\
		&x_{k+1} = A x_k + B K \bs{y}_k(x_k) & \text{(Control deployment)}
	\end{array}
	$$
	\vspace{-.3cm}
\end{algorithm}

Let us now consider the sequence of instructions summarized in Algorithm~\ref{alg:two_layer}. For a given state of the \gls{LTI} system $x_k$, at the first step the agents compute the (unique, in view of Lemma~\ref{lemma:properties}.(i)) \gls{GNE} $\bs{y}^*(x_k)$ through any \gls{GNE} seeking procedure available in the literature. Examples of fully distributed algorithms can be found, for instance, in \cite{belgioioso2018projected,yi2019operator,belgioioso2022distributed}, which are here generically represented by the mapping $\mathsf{GNE} : \R^n \to \R^p$ returning the unique point in $\mc S(\cdot)$, i.e., $\mathsf{GNE}(x)=\mc S(x)$. Once computed $\bs{y}^*(x_k)$, the (linear, in the agents' decisions) controller in \eqref{eq:controller} is then implemented on the \gls{LTI} system.
We thus investigate the co-evolution and the equilibrium of the following interconnected dynamics:
\begin{equation}\label{eq:closed_loop_dynamics}
	x_{k+1} = A x_k + B K \bs{y}^*(x_k), \textrm{ with } \bs{y}^*(x_k) = \mathsf{GNE}(x_k) .
\end{equation}

\begin{remark}
	The implementation of Algorithm~\ref{alg:two_layer} requires a setting consisting of a fast dynamics for the agents taking part to the \normaltext{\gls{GNEP}} in \eqref{eq:single_prob}, and a slow dynamics for the \normaltext{\gls{LTI}} system in \eqref{eq:linear_dynamics}. Note that this is the case if, e.g., \eqref{eq:linear_dynamics} characterizes a certain dynamics over a (possibly large) graph where the information exchange among nodes is dictated by social or physical interactions. In the case study described in \normaltext{\S \ref{sec:case_study}}, for instance, this two time-scale condition is met, since the computation of an equilibrium $\bs y_k = \bs y_k(x_k)$ (i.e., the investment of the firms) to implement the control action through $K \bs y_k$ requires less time than the actual market reaction in terms of product sales (i.e., the followers' state evolution $x_{k+1} = A x_k + B K \bs{y}_k(x_k)$). The energy flexibility market application mentioned in \normaltext{\S \ref{subsec:motivating_examples}} is another example fitting this requirement. In that case, residential prosumers owning and operating a diverse distributed energy resource portfolio should react to a specific energy request made by some local aggregator. The physical quantity of interest, in that case, is the energy accumulated by the aggregator itself that is monitored on a slower time-scale compared to the time required by the agents to find some equilibrium $\bs{y}_k(x_k)$ \normaltext{\cite{heinrich2021local,evangelopoulos2022heterogeneous}}.
	\hfill$\square$
\end{remark}
\begin{remark}\label{rem:state_constraints}
	For given controller gains in \eqref{eq:controller}, in view of the linear dynamics in \eqref{eq:linear_dynamics} we note that state constraints can be equivalently recast as coupling constraints affecting the agents' strategies, and thus included into $\Omega(\cdot)$ directly. In fact, for a given $x \in \mc{X}$, we shall additionally impose $(A x + B K_i y_i + B \sum_{j \in \mc I \setminus \{i\}} K_j y_j) \in \mc{X}$, which amount to linear constraints in the collective vector of strategies (provided that $\mc X$ is).
	\hfill$\square$
\end{remark}

\subsection{Certificates}

%
%
By making use of the quadratic constraint in \eqref{eq:IQC} and performing an algorithmic stability analysis \cite{megretski1997system,lessard2016analysis}, we now derive sufficient conditions certifying that some controller $K$ is able to drive the closed-loop dynamics in \eqref{eq:closed_loop_dynamics}, directly following from Algorithm~\ref{alg:two_layer}, to a co-evolutionary equilibrium: 
\begin{theorem}\label{th:stability}
	Let $\Lambda(A) \subset \mc B_1$, and let the controller gains $K_i \in \R^{m \times p_i}$ in \eqref{eq:controller} be fixed, for all $i \in \mc I$. If there exist a matrix $\blue{X} \in \bbS^n_{\succ 0}$ and coefficients $\blue{\lambda}\geq 0$, $\blue{\rho} \in [0,1)$ so that
	\ifTwoColumn
\else
		\begin{equation}\label{eq:MI}
		\begin{bmatrix} A^\top \blue{X} A-\blue{\rho}^2 \blue{X}  & A^\top \blue{X} B K  \\ (\blue{X} B K)^\top A  & (B K)^\top \blue{X} B K \end{bmatrix} + \blue{\lambda} \begin{bmatrix}(\theta/\eta)^2 I & \phantom{-}\bs 0 \\ \bs 0 & -I \end{bmatrix} \preccurlyeq \bs 0 \,
		\end{equation}
	\fi
	holds true, then the sequence $\{(x_k, \bs{y}_k(x_k))\}_{k \in \N}$ generated by Algorithm~\ref{alg:two_layer} satisfies $(x_k, \bs{y}_k) \in \mc{X} \times \{\Omega(x_k) \cap \mc Y\}$, for all $k \in \N$, and 
	converges exponentially fast to a co-evolutionary equilibrium of the \normaltext{\gls{GNEP}} $\Gamma$ in \eqref{eq:single_prob} and \normaltext{\gls{LTI}} system in \eqref{eq:linear_dynamics}. Specifically, $\lim\limits_{k \to \infty} (x_k, \bs{y}_k(x_k)) = ((I-A)^{-1} B K \bs{y}^*, \bs{y}^*)$.
	\hfill$\square$
\end{theorem}

\begin{remark}
	Depending on the problem at hand, requiring that $\Lambda(A) \subset \mc B_1$ may not be too restrictive -- see the case study in \normaltext{\S \ref{sec:case_study}}. Under some reachability assumption on \eqref{eq:linear_dynamics}, however, one can always find some gain matrix $H \in \R^{m \times n}$ so that $(A + BH) \reqdef \bar{A}$ is Schur. In this case, the controller \eqref{eq:controller} reads as $\kappa(x,\bs{y}) = H x + K \bs{y}(x)$ and the analysis above can be adapted with $\bar{A}$ in place of the matrix $A$.
	\hfill$\square$
\end{remark}

In case the controller $\kappa$ is chosen as in \eqref{eq:controller} for fixed control gains $K_i$, $i \in \mc I$, meeting the condition in \eqref{eq:MI} implies exponential convergence of the sequence generated by Algorithm~\ref{alg:two_layer} to a co-evolutionary equilibrium. Specifically, for a closed-loop system characterized by some quadratic constraint as in \eqref{eq:IQC}, satisfying \eqref{eq:MI} allows us to construct a quadratic function $V(x) \eqdef (x - x^*)^\top X (x - x^*)$, serving as Lyapunov function for the autonomous, nonlinear system \eqref{eq:closed_loop_dynamics}, for which can be proven that $V(x_k) \le \rho^{2k} V(x_0)$. The coefficient $\rho$ then plays the role of the contraction rate of the closed-loop system. 


\subsection{Discussion on the conditions in Theorem~\ref{th:stability}}
Besides providing a mean to certify offline
 the stability and performance of the interconnected system at hand, the matrix inequality in \eqref{eq:MI} however poses few practical challenges. 

We note that, in fact, even for a fixed $K$, the condition in \eqref{eq:MI} is nonlinear in the decision variables $X$, $\lambda$ and $\rho$, and it is therefore nontrivial to find a solution (if one does exist) in a computationally efficient way. This issue however can be mitigated by selecting a pertinent $\rho\in[0,1)$ beforehand, 
and then certifying the existence of a $\rho$-contracting, Lyapunov-like function via the following \gls{LMI}:
\ifTwoColumn
\else
\begin{equation}\label{eq:LMI}
	\begin{bmatrix} A^\top \blue{X} A-{\rho}^2 \blue{X}  & A^\top \blue{X} B K  \\ (\blue{X} B K)^\top A  & (B K)^\top \blue{X} B K \end{bmatrix} +  \blue{\lambda}\begin{bmatrix}(\theta/\eta)^2 I & \phantom{-}\bs 0 \\ \bs 0 & -I \end{bmatrix} \preccurlyeq \bs 0 .
\end{equation}
\fi

For given matrices $K_i$, one could thus check immediately whether the underlying controller is stabilizing for \eqref{eq:closed_loop_dynamics} by solving \eqref{eq:LMI} with a value of $\rho$ close to $1$ (or even equal to $1$ in case marginal stability is a consideration), and then refine it to find the ``best'' contraction rate via, e.g., a bisection method. 
\begin{remark} 
	Both in~\eqref{eq:MI} and \eqref{eq:LMI}, we could fix $\lambda =1$ without loss of generality, since the conditions remain valid for any positive scalar multiplication. This is also true for some of the conditions given later on, even if not explicitly mentioned. 
	\hfill$\square$
\end{remark}

How to choose linear gains $K_i$ is however still unclear. The next section thus aims at shedding light on this crucial point.

\section{On the controller design}\label{sec:controller_design}
Once established sufficient conditions to guarantee that certain controller gains $K_i$, $i \in \mc I$, stabilize the closed-loop system in \eqref{eq:closed_loop_dynamics}, we now move on the computational aspect, i.e., we want
to find $K_i$, $i \in \mc I$, so that \eqref{eq:MI},  or \eqref{eq:LMI}, is satisfied.

For simplicity, in the remainder we set $p_i=m$, $i \in \mc I$, although a generalization including tailored $\bs 0$-blocks is possible -- see, for instance, the discussion on the case study in \S \ref{sec:case_study}.

\subsection{The ``light-touch'' principle for the controller synthesis}
We note first that, in view of $\Lambda(A) \subset \mc B_1$, the two matrix inequalities above can be satisfied with $K=\bs{0}_{m\times mN}$, although one could experience issues related to state constraint satisfaction, i.e., $x_k \in \mc{X}$ may not be guaranteed for all $k \in \N$. 

Moreover, the choice $K=\bs{0}_{m\times mN}$ is also not recommended since the different agents will have no incentives to participate in the resulting competitive game, if at the end their control action is totally nullified. Selecting $K=\bs{0}_{m\times mN}$ amounts to a maximal-intervention choice, whereby we decide to take total control of the competition market and effectively shut it down. On the other side, we could have the no-intervention policy of $K_i=I_m$, when we decide that the market will self-regulate with no external intervention. This choice is known in the economic literature as the \emph{Adam Smith's invisible hand} \cite{rothschild1994adam}. 

Thus, the choice $\|K_i\|\leq C$, with $K_i$ as close as possible to $I_m$, is getting more credit and explored as a middle ground, introducing a possible regulation to an otherwise free market. This type of methodology amounts to the so-called \emph{light-touch} policy. Here $C$ is the maximal amount of incentives that can be given to the participating companies. When $C<1$, the incentives can be effectively seen as taxes that reduce the influence of companies on the state $x$.

The light-touch policy yields the following control problem:
$$
\left\{
	\begin{aligned}
		&\underset{\blue{K}}{\textrm{min}} && \|\blue{K}-(I_m \otimes \bsone_N^\top)\| \\
		& \textrm{ s.t. } && \|\blue{K}_i\|\leq C, \, [\blue{K}_i]_{hk}\geq0, \, \forall i \in \mc I.
	\end{aligned}
\right.
$$

However, introducing these additional requirements and na\"ively solving \eqref{eq:MI} (or \eqref{eq:LMI}) also for $K$ leads to additional nonlinearities, as well as $\theta$ is a function of $K$ itself. Motivated by the considerations above, we define $\bar \theta$ as the parameter characterizing Standing Assumption~\ref{standing:strong_monotonicity}.(ii) when each $\|K_i\| \leq C$, which  hence satisfies $\bar \theta \le \|\col((Q_i)_{i \in \mc I}) A\|\|B\| C$ and enables us to rewrite the quadratic constraint in \eqref{eq:IQC} so that the resulting inequality is immune from the value that $K$ takes. Thus, the following optimization problem generates stabilizing gains $K_i$:
\ifTwoColumn
\else
	\begin{equation}\label{eq:control_opt}
		\left\{
		\begin{aligned}
			&\underset{\blue{K}, \blue{X},\blue{\lambda}}{\textrm{min}} && \|\blue{K}-(I_m \otimes \bsone_N^\top)\|\\
			&\textrm{ s.t. }&&\begin{bmatrix} A^\top \blue{X} A-\rho^2 \blue{X}  & A^\top \blue{X} B \blue{K}  \\ (\blue{X} B \blue{K})^\top A  & (B \blue{K})^\top \blue{X} B \blue{K} \end{bmatrix} + \blue{\lambda} \begin{bmatrix}(\bar{\theta}/\eta)^2 I & \phantom{-}\bs 0 \\ \bs 0 & -I \end{bmatrix} \preccurlyeq \bs 0 ,\\
			&&&\blue{X} \in \bbS^n_{\succ 0}, \blue{\lambda} \ge0, \|\blue{K}_i\|\leq C, [\blue{K}_i]_{hk}\geq0, \, \forall i \in \mc I.
		\end{aligned}
		\right.
	\end{equation}
\fi



By making use of standard continuity arguments one can immediately claim the existence of some small enough gain $K$ so that \eqref{eq:control_opt} enjoys a solution. However, how to derive conditions (or even a convex reformulation of \eqref{eq:control_opt}) under which such a problem can be solved efficiently is not straightforward. 

\subsection{Scalar regulation}

A possible simplification leading to a more tractable program that can be handled by available solvers requires one to scale the action of different agents by the same scalar amount~\footnote{Here we let $C=1$ for simplicity, otherwise we can also pick $\omega \in [0,C]$. }, say $\omega \in [0,1]$, so that the controller in \eqref{eq:controller} happens to coincide with $\kappa(x, \bs y(x)) = \omega \sum_{i \in \mc I} y_i(x) = \omega \, (I_m \otimes \bsone_N^\top) \, \bs y(x)$, i.e., setting $K_i = \omega I_m$ for all $i \in \mc I$. Looking at the case study detailed in \S \ref{sec:case_study}, this approach is meant to reflect a so-called light-touch regulation dictated for instance by anti-trust reasons or protecting competition. We have the following result:
\begin{proposition}\label{prop:controller-omega-BMI}
	Let $\kappa(x, \bs y) = \omega \sum_{i \in \mc I} y_i(x)$ and $\rho\in[0,1)$. Then, by defining $\bar B \eqdef B \otimes \bsone_N^\top$, and $\hat \theta \le \|\col((Q_i)_{i \in \mc I}) A\|\|B\|$, \eqref{eq:control_opt} reduces to the following \normaltext{\gls{BMI}}:
	\begin{equation}\label{eq:controller-omega-BMI}
	\left\{
	\begin{aligned}
		&\underset{\blue{\omega}, \blue{X},\blue{\lambda}}{\normaltext{\textrm{min}}} && -\blue{\omega}\\
		&~\normaltext{\textrm{ s.t. }}&&\begin{bmatrix} -{\rho}^2 \blue{X} +\blue{\lambda}(\hat{\theta}/\eta)^2 \blue{\omega}^2 I & \phantom{-}\bs 0 & A^\top \blue{X}    \\ \bs 0 & -\blue{\lambda}I & \blue{\omega} \bar B^\top \blue{X} \\ \blue{X} A & \blue{\omega} \blue{X} \bar B & - \blue{X} \end{bmatrix} \preccurlyeq \bs 0 ,\\
		&&& \blue{\omega}\in [0,1] , \blue{\lambda} \ge0,  \blue{X} \in \bbS^n_{\succ 0}.
	\end{aligned}
	\right.
	\end{equation}
%
	\hfill$\square$
\end{proposition}

%

%

\begin{algorithm}[!t]
	\caption{Bisection-like approach to solve \eqref{eq:controller-omega-BMI}}\label{alg:bisection}
	\DontPrintSemicolon
	\SetArgSty{}
	\smallskip
	\textbf{Initialization:} Choose $\varepsilon > 0$, $\varsigma \in (0,1)$, set $t = 0$, $\rho_0 = \varepsilon$, $\omega_0 = 1$, \texttt{flag} $= 0$\\
	\smallskip
	\While{\texttt{flag} $= 0$}{
		\smallskip
		Solve \gls{LMI}:
		\begin{equation}\label{eq:LMI_test}
			\left\{
			\begin{aligned}
				&\begin{bmatrix} -\rho_t^2 \blue{X} +\blue{\lambda}(\hat{\theta}/\eta)^2 {\omega_t}^2 I & \phantom{-}\bs 0 & A^\top \blue{X}    \\ \bs 0 & -\blue{\lambda}I & \omega_t \bar B^\top \blue{X} \\ \blue{X} A & \omega_t \blue{X} \bar B & - \blue{X} \end{bmatrix} \preccurlyeq \bs 0 \\ &\blue{X} \in \bbS^n_{\succ 0}, \blue{\lambda} \ge0
			\end{aligned}
			\right.
		\end{equation}
		
		\uIf{\eqref{eq:LMI_test} \texttt{infeasible}}{
			\uIf{$\omega_{t} \le \varepsilon$}{
				\smallskip
				Update $\omega_{t+1} = 1$, $\rho_{t+1} = \textrm{min}\{\rho_t + \varsigma, 1\}$\\
				\smallskip
			}
			\Else{
				\smallskip
				Update $\omega_{t+1} = \textrm{max}\{\omega_{t} - \varsigma, \varepsilon\}$, $\rho_{t+1} = \rho_t$ \\
				\smallskip
			}
		}
		\Else{
			\texttt{flag} $= 1$
		}
		\smallskip
		Set $t = t+1$
	}
\end{algorithm}

In case solvers to compute a solution to \eqref{eq:controller-omega-BMI} are not available, 
one could also devise a bisection-like procedure, as the one in Algorithm~\ref{alg:bisection}, to find a suitable matrix $X$ by fixing $\rho$ and $\omega$ iteratively so that the \gls{BMI} in \eqref{eq:controller-omega-BMI} actually reduces to an \gls{LMI}. 

Bearing in mind that a desirable solution seeks for a scaling factor $\omega$ guaranteeing the least intervention possible (i.e., $\omega$ close to one) with the best closed-loop performance (i.e., the smallest $\rho$ possible), Algorithm~\ref{alg:bisection} requires one to initialize $\rho$ with some small $\varepsilon > 0$ and  $\omega = 1$, and then solve the \gls{LMI} described in \eqref{eq:LMI_test}, resulting from \eqref{eq:controller-omega-BMI}. In case this latter has no solution, the scaling factor $\omega$ is then reduced by some predefined quantity $\varsigma \in (0,1)$, while keeping $\rho$ fixed. This latter is increased by, e.g., the same $\varsigma$, only if a solution to \eqref{eq:LMI_test} is not found with a large enough value of $\omega$ (e.g., the same $\varepsilon$ or a higher value). In this way, Algorithm~\ref{alg:bisection} stops when a solution to \eqref{eq:LMI_test} exists with the ``largest'' value of $\omega$ and the ``smallest'' of $\rho$. If \eqref{eq:LMI_test} has no solution with $\omega = \varepsilon$ and $\rho = 1$, however, according to Theorem~\ref{th:stability} the nonlinear controller $\kappa(x, \bs y(x)) = \omega \, (I_m \otimes \bsone_N^\top) \,\bs y(x)$ is not theoretically guaranteed to stabilize the co-evolution in \eqref{eq:closed_loop_dynamics}, though it could still behave well in practice as condition \eqref{eq:MI} is only sufficient.

Finally, while via Algorithm~\ref{alg:bisection} one can select one ``optimal'' pair $(\omega, \rho)$, nobody prevents us to look for all the pairs $(\omega, \rho)$ for which \eqref{eq:LMI_test} is verified. This leads to explicit trade-offs between regulation and reactivity of the competitive market.

\section{Case study: Advertising through influencers with digital regulation}\label{sec:case_study}

We now elaborate on the motivating example introduced in \S \ref{subsec:motivating_examples} and how it fits the proposed framework. We first characterize some technical properties of the model adopted, then devise a dimension-reduction procedure to make the resulting \glspl{BMI} verification computationally appealing, and finally conduct numerical simulations to corroborate our results.

\subsection{Revising the problem and mathematical model}
Refer to Fig.~\ref{cool.pic} and consider $N$ firms, $m$ influencers, and $n_F$ followers/customers.  We assume each firm $i \in\mc I=\{1,\ldots,N\}$ wants to solve an inter-dependent optimization problem as:
\begin{equation}\label{eq:single_prob_example_schematic}
		\forall i \in \mc I : \left\{
		\begin{aligned}
			& \underset{y_i \in \mc{Y}_i}{\textrm{min}} && \frac{1}{2}\left(\|A_F x_F + B_F u(\bs y) - \bar{x}^i_F \|^2_{Q_i} + \|y_i\|^2_{R_i}\right)\\
			& \hspace{.1cm}\textrm{ s.t. } && C_i y_i + \textstyle\sum_{j \in \mc I \setminus \{i\}} C_j y_j \le d,
		\end{aligned}
		\right.
\end{equation}
which models the selfish interests of $N$ firms, which in our framework coincide with the agents taking part to the Nash game in \eqref{eq:single_prob}. These latter, indeed, aim at selling a desirable quantity of products $\bar{x}^i_F \geq 0$ guaranteeing a certain degree of profit, and hence invest their money $y_i \in \mc Y_i \subset \R^m$ to pay $m$ influencers in order to advertise them (each $\mc Y_i$ limits the available budget). Social influencers, on their side, are connected through, e.g., social networks, with the population of consumers via matrices $(A_F, B_F)$, and hence can steer the sale of those products throughout the network. The shared constraints with $C_i \in \R^{l \times p_i}$ and $d \in \R^l$ may reflect possible income limitations the social influencers have to deal with, while $\mc X$ may represent production limitations, shortages or third party restrictions. Here $Q_i, R_i$ are positive definite weight matrices, and as before $\bs{y} \eqdef \col((y_i)_{i \in \mc I})$.

Note that \eqref{eq:single_prob_example_schematic} captures the selfish nature of each firm. The cost function combines the willingness of companies to achieve their selling goal (first term), while trying to pay influencers as little as possible (second term). Notably, the first term is coupled with the customers' dynamics, for which $u$ represents the control input and is a function of the resources $\bs y$.

We will now look at the system matrices $A_F, B_F$ and see how to model them in our case. The system consisting of influencers and potential consumers (i.e., their followers) can be abstracted as a static network of $M$ agents in total that locally exchange information according to a connected and undirected graph $\mc{G} \eqdef (\mc{M}, \mc{E}, w)$ with known topology, $\mc{M} \eqdef \{1, \ldots, M\}$ and $\mc{E} \eqdef \set{(i, j)}{i, j \in \mc{M}, i \neq j}$. Set $\mc{M}$ indexes the agents, which for simplicity are assumed to be associated with a scalar variable $x_i \in \R$ (the extension to a vector is straightforward), $\mc{E}$ denotes the information flow links dictated by the social network, and $w \in \R_{\ge 0}^{|\mc{E}|}$ the weights on the edges reflecting the actual influence for the considered social network.
Then, we consider an instance where the population of consumers follows a weighted agreement protocol that is also affected by external inputs $u \in \R^m$ injected at $m$ specific nodes represented by the influencers. We can thus split the set $\mc{M} = \mc{M}_F \cup \mc{M}_I$ into followers ($\mc{M}_F$, $n_F \eqdef |\mc{M}_F|$) and influencer nodes ($\mc{M}_I$, $|\mc{M}_I| = m$) so that the dynamics for each $i \in \mc{M}_F$ reads as:
\begin{equation}\label{eq:LTI_network_single_agent}
x_i^+ = \alpha_i x_i + \tau \sum_{j \in \mc N_i \cap  \mc{M}_F} w_{i,j}(x_j - x_i) + \tau\!\!\sum_{h \in \mc N_i \cap  \mc{M}_I} w_{i,h} (x_h - x_i),
\end{equation}
where each $\alpha_i\in(0,1]$ denotes a \emph{susceptibility to persuasion}-like term, reflecting standard Friedkin-Johnsen models \cite{friedkin1999social}.
Given their specific role, the influencer nodes hence affect the followers' dynamics through ``directed edges'' in the sense that they do no not follow any local, agreement-like protocol whose control contribution is assigned through weights $w_{i,h}$.
In accordance with the splitting of the nodes $\mc{M} = \mc{M}_F \cup \mc{M}_I$, the weighted incidence matrix $D \in \R^{n \times |\mc{E}|}$ characterizing $\mc{G}$ can also be partitioned as $D = \col(D_F, D_I)$, with $D_F \in \R^{n_F \times |\mc{E}|}$ and $D_I \in \R^{m \times |\mc{E}|}$, thus leading to the following \gls{LTI} dynamics characterizing the followers' states $x_F \eqdef \col((x_i)_{i \in \mc{M}_F})$ \cite{mesbahi2010graph}
\begin{equation}\label{eq:LTI_network}
	x_F^+ = A_F x_F + B_F u ,
\end{equation}
where $A_F \eqdef A_F(w) = \diag((\alpha_i)_{i\in\mc{M}_F}) - \tau D_F W D_F^\top$, $B_F \eqdef B_F(w) = - \tau D_F W D_I^\top$, $u \eqdef \col((x_i)_{i \in \mc{M}_I})$ and $W \eqdef \diag(w) \in \R^{|\mc{E}| \times |\mc{E}|}$, and sampling time $\tau > 0$ to be suitably determined according to the following result:
\begin{proposition}\label{prop:stab}
	Let $\mc{G}$ be a connected and undirected graph, $W \succ 0$ and $\alpha_i \in (0,1]$, for all $i\in\mc M_F$. Then, $A_F$ is a symmetric matrix and if $\tau \in (0, \mathrm{min}_{i\in\mc M_F}\{1+\alpha_i\}/\lambda_\normaltext{\textrm{max}}(L_F))$, $\Lambda(A_F) \subset \mc B_1$, where $L_F \eqdef D_F W D_F^\top$. 
	\hfill$\square$
\end{proposition}
Then, choosing a small enough sampling time $\tau$ for the \gls{LTI} dynamics in \eqref{eq:LTI_network}, interconnected with the \gls{GNEP} in \eqref{eq:single_prob_example_schematic}, allows one to meet the condition in Theorem~\ref{th:stability}, thus making the problem suitable to be analysed with the tools developed. 


The remuneration process involving companies and influencers, however, can not be arbitrary. Some works, indeed, have recently investigated how to regulate such digital markets from a legislation perspective \cite{stewart2020trouble,OECD2021ex,EU2022digital,goanta2020regulation}. Therefore, 
since influencers have to declare their revenues and conflict of interests, it seems reasonable to assume that a government or some third party is allowed to charge (or eventually incentivize in case it wants to steer the public opinion as well) influencers and/or advertisements through the gain matrices $K_i$, $i \in \mc I$, according to choice of the control input $u(\cdot)$ made in \eqref{eq:controller}.

\subsection{A dimension-reduction approach}\label{subsec:dim_red}


We now develop a dimension-reduction procedure for the resulting control design problem. While the \gls{BMI}~\eqref{eq:controller-omega-BMI} could be applied directly here to devise a light-touch regulation $\omega$,  matrices $(A_F, B_F)$ may be of very high dimension (in case of thousands, or even millions of followers). This intrinsically hinders the practical solvability~of~\eqref{eq:controller-omega-BMI}. 
We can however circumvent this issue at the expense of introducing some conservatism, and deriving a condition whose size is independent on the number of followers $n_F$, influencers $m$, and companies $N$. To do that, we will make use of a standard full-block $S$-procedure, as well as tools from~\cite{Massioni2014,DePasquale2020}. 

Consider the dynamical system~\eqref{eq:LTI_network}. In view of the light-touch principle and resulting controller structure described in \S \ref{sec:controller_design}, which will also be adopted here to steer the behavior of the population of consumers, we define $\bar B_F \eqdef B_F  \otimes \bsone_N^\top$ so that the nonlinear control law will amount to $u=\omega \bs y(x_F) \in \R^{mN}$. 
\begin{remark}
	Even though we focus on the controller structure derived in \normaltext{\S \ref{sec:controller_design}}, the mathematical developments given next also hold true for more general controllers as in \eqref{eq:controller}.
	\hfill$\square$
\end{remark}
From now on, we thus focus on the dynamics:
\begin{equation}\label{eq:LTI_network_aug}
	x_F^+ = A_F x_F + \bar B_F u,
\end{equation}
and we will assume that $\alpha_i = \alpha$ for all $i\in\mc M_F$. This latter assumption could be extended by considering, e.g., \cite{Massioni2014}. 

In addition, it is reasonable to assume here that $m \ll n_F$, and hence without loss of generality we can augment the column space of $B_F$ to be of the same dimension $n_F$ of the state $x_F$ by adding $n_F-m$ virtual influencer nodes with $w_{i,h}=0$ in \eqref{eq:LTI_network_single_agent}. This yields $\bar B_F \in \R^{n_F \times n_F N}$, as well as $u \in \R^{n_F N}$. 

Thus, the introduction of two additional signals $\gamma \in \R^{2n_F}$ and $\zeta \in \R^{n'}$, $n' \eqdef n_F(N+1)$, allows us to rewrite \eqref{eq:LTI_network_aug} as
\ifTwoColumn
\else
	\begin{equation}\label{eq:LTI_Delta}
		\left\{
		\begin{aligned}
			&x_F^+ =  I_{n_F} x_F + [I_{n_F} \; I_{n_F}] \gamma \reqdef \mathcal{A} x_F + \mathcal{B} \gamma,\\
			&\zeta= [I_{n_F} \; \bs 0_{n_F \times n_F N}]^\top x_F + [\bs 0_{n_F N \times n_F} \; I_{n_F N}]^\top u \reqdef \mathcal{C} x_F + \mathcal{D} u,\\
			&\gamma = \begin{bmatrix} -\tau D_F W D_F^\top & \bs 0_{n_F \times n_F N} \\ \phantom{-}\bs 0_{n_F \times n_F } & \bar B_F \end{bmatrix} \zeta = \reqdef \Delta \zeta,
		\end{aligned}
		\right.
	\end{equation}
\fi
with topology-dependent, dense matrices $\tau L_F$ and $\bar B_F$.

Putting temporarily aside the controller synthesis, i.e., the tuning of the scaling factor $\omega \in [0,1]$, set equal to one for the moment, we discuss next the closed-loop stability of the dynamical system \eqref{eq:LTI_Delta} with $u=\bs y(x_F)$. In what follows we indicate with $\star$ the matrix that post-multiply the square one in the middle, e.g., $(\star)^\top A_F V = V^\top A_F V$, for some $V \in \R^{n_F \times r}$.

\begin{theorem}\label{th:CL_S_proc} 
	Let $\tau \in (0, (1+\alpha)/\lambda_\normaltext{\textrm{max}}(L_F))$. If there exist matrices $\blue{X} \in \bbS_{\succ 0}^{n_F}$, $\blue{R}\in\bbS^{n'}$, $\blue{T}\in\bbS^{2n_F}$, $\blue{S}\in\mathbb{R}^{n' \times 2n_F}$, and coefficients $\blue{\lambda}\geq 0$, $\blue{\rho} \in [0,1)$ so that
	\ifTwoColumn
\else	
	\begin{align}
		& (\star)^\top \begin{bmatrix} \blue R & \blue S \\ \blue S^\top & \blue T \end{bmatrix}\begin{bmatrix} I_{n'} \\ \Delta\end{bmatrix} \succ \bs 0, \text{ and } \label{start.lmi}\\
		&
		(\star)^\top \left[\begin{array}{c|c} \begin{array}{cc} \blue{X} & \phantom{-}\bs0 \\ \bs0 & -\blue{X}\end{array} & {\bf 0}\\ \hline {\bf 0} & \begin{array}{cc} \blue R & \blue S \\ \blue S^\top & \blue T\end{array}  
		\end{array}\right] \left[\begin{array}{ccc} \mathcal{A}  & \mathcal{B} & \bs 0   \\ 
			\blue\rho I_{n_F} &  \bs 0 & \bs 0\\ \hline
			\mathcal{C} & \bs 0 & \mathcal{D} \\
			\bs 0 & I_{2n_F} & \bs 0
		\end{array}\right] +\blue{\lambda} \begin{bmatrix} (\bar\theta/\eta)^2 I_{n_F} & \bs 0 & \bs 0\\ \bs 0 & \bs 0 & \bs 0 \\\bs 0 & \bs 0& -I_{n_F N} 
		\end{bmatrix} \preccurlyeq \bs 0, \label{end.lmi}
	\end{align}
	\fi
	hold true, then the sequence $\{(x_{F,k}, \bs{y}_k(x_{F,k}))\}_{k \in \N}$ generated by Algorithm~\ref{alg:two_layer} satisfies $(x_{F,k}, \bs{y}_k) \in \mc{X} \times \{\Omega(x_{F,k}) \cap \mc Y\}$, for all $k \in \N$, and 
	converges at an exponential rate to a co-evolutionary equilibrium of the \normaltext{\gls{GNEP}} $\Gamma$ in \eqref{eq:single_prob_example_schematic} and \normaltext{\gls{LTI}} system in \eqref{eq:LTI_network_aug}, i.e., $\lim\limits_{k \to \infty} (x_{F,k}, \bs{y}_k(x_{F,k})) = ((I-A_F)^{-1} \bar B_F \bs{y}^*, \bs{y}^*)$.~\hfill$\square$
\end{theorem}


To obtain the desired dimensionality reduction from the stability conditions just derived, we consider now only scalar (or reduced dimension) decision variables and multipliers, as well as we set $S =\bs0_{n' \times 2n_F}$. This is a common practice for dimensionality reduction, which however introduces some degree of conservatism. In particular, we will consider
$X = \chi I_{n_F}$, $R = \diag(r_1,r_2 I_N)\otimes I_{n_F}$, $T = \diag(t_1,t_2) \otimes I_{n_F}$, with $\chi > 0$, $r_j, t_j \in \mathbb{R}$, $j=1, 2$. We have the following result:

\begin{theorem}\label{th:decomposition}
%
Let $\delta_{\normaltext{\textrm{max}},1}$ be the maximum singular value of $-\tau D_F W D_F^\top$, and $\delta_{\normaltext{\textrm{max}},2}$ that of $\bar B_F$, let $\tau \in (0, (1+\alpha)/\lambda_\normaltext{\textrm{max}}(L_F))$. By setting $X = \blue{\chi} I_{n_F}$, $R = \diag(\blue{r_1}, \blue{r_2} I_N)\otimes I_{n_F}$, $T = \diag(\blue{t_1}, \blue{t_2}) \otimes I_{n_F}$, and $S =\bs0_{n' \times 2n_F}$, the statement in Theorem~\ref{th:CL_S_proc} holds true in case there exist scalars $\blue{\chi} > 0$, $\blue{\lambda} \geq 0$, $\blue{r_i} >0$, $j = 1, 2$, and $\blue{\rho} \in [0,1)$ so that:
\ifTwoColumn
\else
\begin{align}
	&(\star)^\top \begin{bmatrix} \blue{r_j} & 0 \\ 0 & \blue{t_j} \end{bmatrix}\begin{bmatrix} 1 \\ \delta_{\normaltext{\textrm{max}},j}\end{bmatrix} > 0,~j = 1,2, \text{ and } \label{start.lmi2}\\
	&(\star)^\top \!\!\left[\begin{array}{c|c} \begin{array}{cc} \blue{\chi} & \phantom{-}0 \\ 0 & -\blue{\chi}\end{array} & {\bf 0}\\ \hline {\bf 0} & \begin{array}{cc} \blue{r} & \bs 0 \\ \bs 0 & \blue{t}\end{array}  
	\end{array}\right] \!\!\left[\begin{array}{ccc} \alpha  &\!\! \bsone_2^\top &\!\! 0   \\ 
		\blue{\rho}  &\!\!  \bs 0 &\!\! 0\\ \hline
		[1 \; 0]^\top &\!\! \bs 0 &\!\! [0 \; 1]^\top \\
		\bs 0 &\!\! I_{2} &\!\! \bs 0
	\end{array}\right] + \blue{\lambda} \begin{bmatrix} (\hat\theta/\eta)^2 & \bs 0 & \phantom{-}\bs 0\\ \bs 0 & \bs 0 & \phantom{-}\bs 0 \\ \bs 0 & \bs 0& -1  
	\end{bmatrix} \preccurlyeq \bs0, \label{end.lmi2}
\end{align}
\fi
where $\blue{r} \eqdef \diag(\blue{r_1},\blue{r_2})$ and $\blue{t} \eqdef \diag(\blue{t_1},\blue{t_2})$.
\hfill$\square$
\end{theorem}

Specializing \eqref{start.lmi}--\eqref{end.lmi}, the conditions reported in Theorem~\ref{th:decomposition} allow one to handle a potentially large number of followers, as they characterize the co-evolution of the \normaltext{\gls{GNEP}} $\Gamma$ in \eqref{eq:single_prob_example_schematic} and \normaltext{\gls{LTI}} system in \eqref{eq:LTI_network_aug}. In addition, we note that the presented dimension-reduction framework enables us to consider also time-varying weights $W$, links set $\mc E$, or uncertainties affecting the followers' dynamics. As long as we are able to compute the maximal singular value of $\Delta$ (or estimates an its upper bound), indeed, conditions \eqref{start.lmi2}--\eqref{end.lmi2} can still be verified and, albeit more conservative, they allow one to cover relevant extensions to the case study described here.

The design of a light-touch controller $\omega \in [0,1]$ in the spirit of \S \ref{sec:controller_design} can now be done by considering $\omega \mc{D}$ instead of just $\mc D$, and slightly modifying the condition in \eqref{end.lmi2} to obtain:
\begin{equation}\label{eq:lmi2_controller}
\begin{bmatrix} (\alpha^2-\blue{\rho}^2) \blue{\chi}  + \blue{r_1} + \blue{\lambda} (\hat{\theta}/\eta)^2 \blue{\omega}^2&  \alpha\blue{\chi} \bsone^\top_2 & 0 \\ 
\alpha \blue{\chi} \bsone_2  & \blue{\chi} I_2 + \blue{t} & \bs0 \\ 0 & \bs0 & \blue{\omega}^2 \blue{r_2} - \blue{\lambda}
\end{bmatrix} \preccurlyeq \bs0.
\end{equation}
Together with \eqref{start.lmi2}, this latter relation can then be solved directly by bisection on $\omega^2$, thus applying exactly the same reasoning of \S \ref{sec:controller_design} and resulting Algorithm~\ref{alg:bisection}.


\subsection{Numerical results}
\begin{table*}[!t]
	\caption{Comparison between original and dimension-reduction approach -- varying the number of followers}\label{tab:procedures_fol}
	\centering
	\begin{tabular}{cccccccccccccccc}
		\toprule
		\multirow{2}[2]{*}{Control design} & \multicolumn{3}{c}{$n_F=50$} & & \multicolumn{3}{c}{$n_F=100$} & & \multicolumn{3}{c}{$n_F=200$} & & \multicolumn{3}{c}{$n_F=1000$}\\
		\cmidrule{2-4} \cmidrule{6-8} \cmidrule{10-12} \cmidrule{14-16} & \multicolumn{1}{c}{\footnotesize{CPU time}} & \multicolumn{1}{c}{$\omega$} & \multicolumn{1}{c}{$\rho$}  &  & \multicolumn{1}{c}{\footnotesize{CPU time}} & \multicolumn{1}{c}{$\omega$} & \multicolumn{1}{c}{$\rho$} & & \multicolumn{1}{c}{\footnotesize{CPU time}} & \multicolumn{1}{c}{$\omega$} & \multicolumn{1}{c}{$\rho$} & & \multicolumn{1}{c}{\footnotesize{CPU time}} & \multicolumn{1}{c}{$\omega$} & \multicolumn{1}{c}{$\rho$}\\
		\midrule
		\eqref{eq:controller-omega-BMI} & 12.5~[s] & 0.96 & 0.87 &  & 681.8~[s] & 0.92 & 0.88 & & > 3600~[s] & * & * & & * & * & *\\
		\midrule
		\eqref{start.lmi2} + \eqref{eq:lmi2_controller} & 0.25~[s] & 0.92 & 0.86 &  &0.13~[s] &0.92 & 0.82 & &0.14~[s] &0.92& 0.83 & & 3.24~[s] & 0.91 & 0.83\\
		\bottomrule
	\end{tabular}
\end{table*}

\begin{table*}[!t]
	\caption{Comparison between original and dimension-reduction approach -- varying the number of influencers}\label{tab:procedures_inf}
	\centering
	\begin{tabular}{cccccccccccccccc}
		\toprule
		\multirow{2}[2]{*}{Control design} & \multicolumn{3}{c}{$m=1$} & & \multicolumn{3}{c}{$m=5$} & & \multicolumn{3}{c}{$m=10$} & & \multicolumn{3}{c}{$m=20$}\\
		\cmidrule{2-4} \cmidrule{6-8} \cmidrule{10-12} \cmidrule{14-16} & \multicolumn{1}{c}{\footnotesize{CPU time}} & \multicolumn{1}{c}{$\omega$} & \multicolumn{1}{c}{$\rho$}  &  & \multicolumn{1}{c}{\footnotesize{CPU time}} & \multicolumn{1}{c}{$\omega$} & \multicolumn{1}{c}{$\rho$} & & \multicolumn{1}{c}{\footnotesize{CPU time}} & \multicolumn{1}{c}{$\omega$} & \multicolumn{1}{c}{$\rho$} & & \multicolumn{1}{c}{\footnotesize{CPU time}} & \multicolumn{1}{c}{$\omega$} & \multicolumn{1}{c}{$\rho$}\\
		\midrule
		\eqref{eq:controller-omega-BMI} & 209.4~[s] & 0.95 & 0.9 &  & 665.2~[s] & 0.95 & 0.89 & & 1430~[s] & 0.95 & 0.88 & & 3254.6~[s] & 0.96 & 0.93\\
		\midrule
		\eqref{start.lmi2} + \eqref{eq:lmi2_controller} & 0.37~[s] & 0.92 & 0.88 &  &0.12~[s] & 0.92 & 0.93 & &0.12~[s] & 0.93 & 0.85 & & 0.11~[s] & 0.92 & 0.80\\
		\bottomrule
	\end{tabular}
\end{table*}
%
%
%
%
%
%
%

%
%

We now implement the closed-loop dynamics in \eqref{eq:closed_loop_dynamics} with light-touch controller designed in \S \ref{sec:controller_design} by solving \eqref{eq:controller-omega-BMI} numerically according to the method presented in Algorithm~\ref{alg:bisection}.

All simulations are run in MATLAB on a laptop with an Apple M2 chip featuring an 8-core CPU and 16 Gb RAM. The obtained matrix inequalities are 
solved with SeDuMi \cite{sturm1999using}.

Specifically, given $n_F$ followers and $m$ influencers, for the noncooperative game among $N$ companies we set $R_i \sim \mc U(1,2) I_m$, $Q_i \sim \mc U(0.001, 0.1) I_{n_F}$, while each $\bar x_F^i = (p_i/n_F) \bsone_{n_F}$ is chosen according to the production power $p_i \sim \mc U(50,500) n_F$ of each company. Local constraints $\mc Y_i$ limit the budget each firm may spend to get its goods advertised by influencers. In particular, an upper bound on the total budget is defined as the product of three quantities: the production power $p_i$, the price per unit $\upsilon_i \sim \mc U(1.8,2.25)$, and the percentage of the proceeds that goes to the influencers $\varrho_i \sim \mc U(0.02,0.08)$. In our case study, we have assumed four types of influencers in accordance to the number of followers they have connection with: small ($n_F/10$), regular ($n_F/5$), rising ($n_F/2$) and macro ($n_F$), with corresponding weights $w_{i,h}$ on the dynamics \eqref{eq:LTI_network_single_agent} of $1.2$, $2.5$, $7.5$ and $12$, respectively. On the other hand, we assume the followers have identical mutual influence to each other, i.e., $w_{i,j}=1$. In addition, each influencer type yields coupling constraints among companies according to the different income limitations the influencers incur on. Specifically, we impose that, for each $j \in \mc M_I$, $\sum_{i \in \mc N} y_i^j \le \iota^j$, where $y_i^j$ is the $j$-th component of decision vector $y_i$ and $\iota^j$ represents the income limitation of influencer $j$, with $\iota^j \sim \mc U(400,2000)$. Finally, we impose an upper bound on the state $\bar x_F$ so that $x_{F,k} \in [\bs 0_{n_F}, \bar x_F]$, with $\bar x_F = (\sum_{i \in \mc N} \bar x_F^{i,1}) \bsone_{n_F}$, to account for shortages of production ($\bar x_F^{i,1}$ is the first element of $\bar x_F^i$).

\begin{figure}[!t]
	\centering
	\ifTwoColumn
\else
	\includegraphics[width=.6\columnwidth]{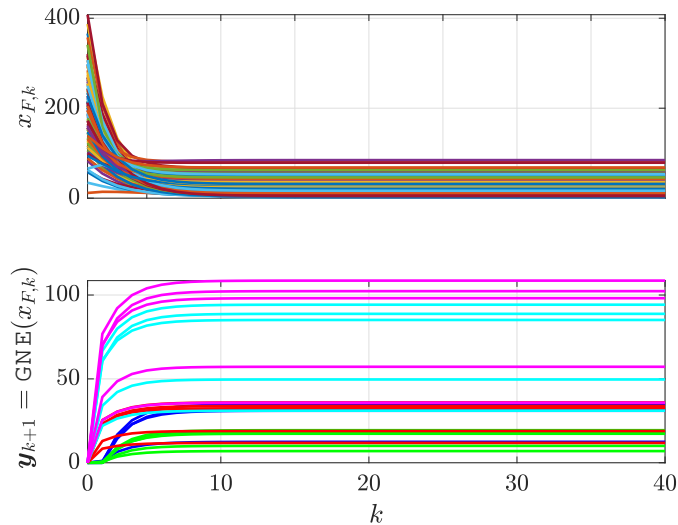}
	\fi
	\caption{Followers' dynamics $x_{F,k}$ in \eqref{eq:LTI_network} and companies' collective decision vector $\bs y_k$ in \eqref{eq:single_prob_example_schematic} co-evolution.}
	\label{fig:co_evol}
\end{figure}

\begin{figure}[!t]
	\centering
	\ifTwoColumn
\else
	\includegraphics[width=.6\columnwidth]{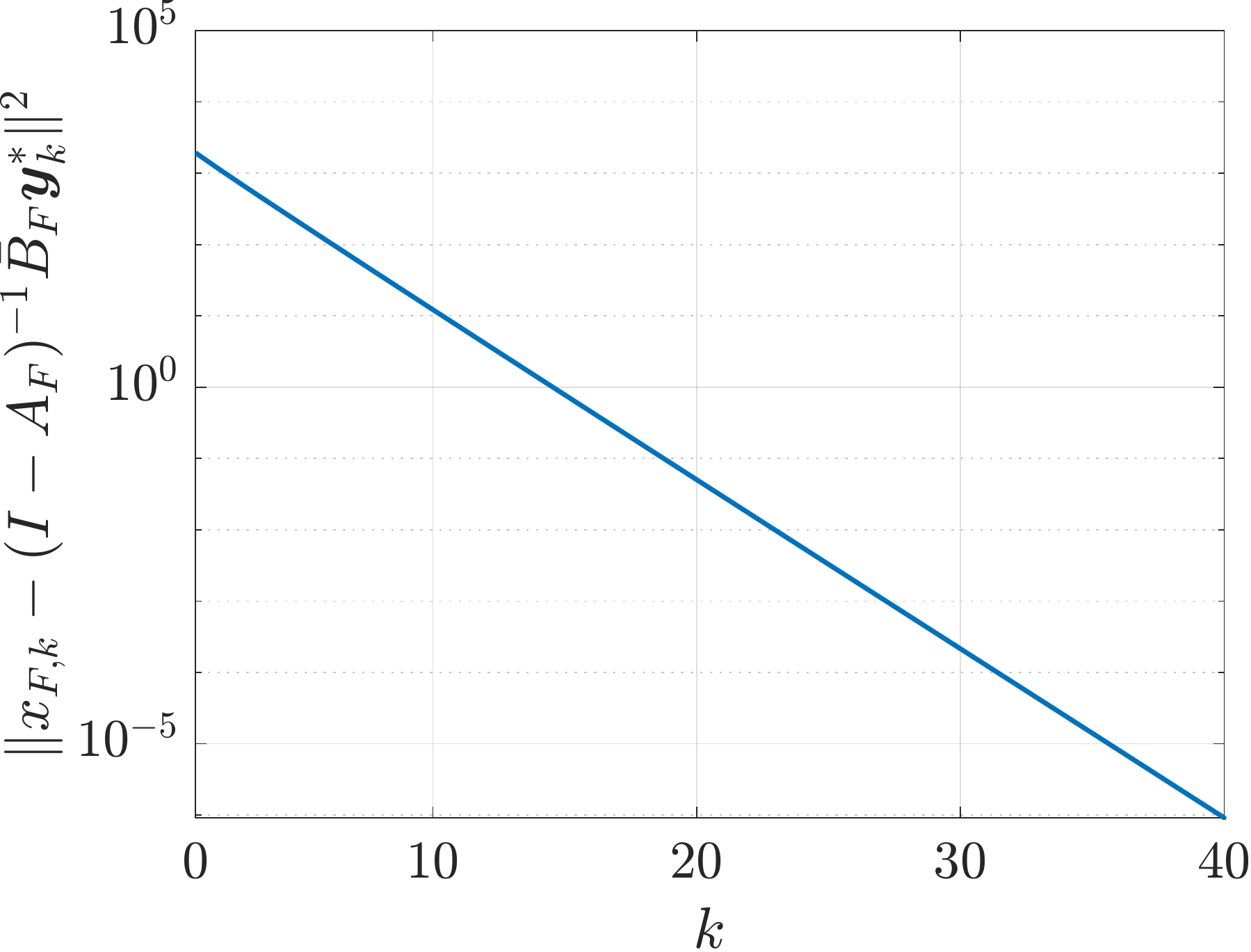}
	\fi
	\caption{Linear convergence to an equilibrium of the co-evolution dynamics driven by Algorithm~\ref{alg:two_layer}. 
	}
	\label{fig:convergence}
\end{figure}

Figures~\ref{fig:co_evol} and \ref{fig:convergence} illustrates the (very fast) co-evolution originating from the \gls{GNEP} in \eqref{eq:single_prob_example_schematic} and dynamics \eqref{eq:LTI_network} for a random instance with $N=10$, $m=5$, $n_F = 100$ and $|\mc E| =582$. The edges underlying the followers' dynamics are randomly generated so that the resulting graph is connected to meet the conditions in Proposition~\ref{prop:stab}. In addition, we have chosen a susceptibility to persuasion $\alpha=0.75$ and a sampling time $\tau=1.75/\lambda_\normaltext{\textrm{max}}(L_F)$.
In particular, from Fig.~\ref{fig:convergence} we appreciate the linear convergence following from Theorem~\ref{th:stability} with light-touch controller $\omega=1$ and $\rho=0.65$ obtained by solving \eqref{eq:controller-omega-BMI} through Algorithm~\ref{alg:bisection} with $\varepsilon=\zeta=0.01$.
The mapping $\mathsf{GNE}(\cdot)$ required to implement Algorithm~\ref{alg:two_layer} coincides with the extragradient method presented in \cite{solodov1996modified}.

We finally compare the original approach to design light-touch controllers presented in \S \ref{sec:controller_design} with the dimension-reduction one of \S \ref{subsec:dim_red}, both solved through the bisection-like method in Algorithm~\ref{alg:bisection}. Specifically, Tables~\ref{tab:procedures_fol} and \ref{tab:procedures_inf} contrast them in terms of CPU time required to find a solution (in seconds) and control performance, i.e., reporting the obtained values for $\omega$ and $\rho$, averaged over $10$  numerical instances for each case.

In particular, Table~\ref{tab:procedures_fol} considers several values for $n_F$, while we use $|\mc E| = 4n_F$, $m=10$ influencers and $N=10$ companies for each example. 
As expected, the control approach based on the solution to \eqref{eq:controller-omega-BMI} is not viable as the dimension of the considered graph grows, while the dimension-reduction procedure obtained by combining \eqref{start.lmi2} and \eqref{eq:lmi2_controller} still makes possible the design of an light-touch controller with far less offline computation.
In fact, the columns referring to the original procedure \eqref{eq:controller-omega-BMI} show that we can obtain a solution in less than 3600 [s] only for $n_F \le 100$, while for $n_F = 200$ simulation was aborted after one hour. When $n_F = 1000$, instead, the solver even crashes.

On the other hand, Table~\ref{tab:procedures_inf} fixes the number of followers $n_F$ to $100$ and considers several values for $m$. The values for $N$ and $|\mc E|$, instead, remain the same as for Table~\ref{tab:procedures_fol}. 
Overall, from our numerical experience on this case study, it seems that the dimension-reduction procedure only produce a minor performance degradation, while requiring significantly less computational costs to find a feasible control solution.

\section{Conclusion}
Motivated by a relevant contemporary application in digital market regulation, we have analyzed the co-evolution arising when the decisions of a population of selfish agents are tightly coupled with an external dynamics. After providing stability results for the closed-loop system, we have established suitable, matrix inequality-based, procedures to design stabilizing controllers, here interpreted as light-touch incentives to steer such an external dynamics while maintaining a certain flavour of tractability in solving the resulting optimization problems. Once developed a mathematical model for an advertising-through-influencers problem with digital regulation, we have additionally devised a dimension-reduction approach to reduce the computational costs required by our procedure. 

\appendix

\textit{Proof of Lemma~\ref{lemma:properties}}: Both results follow from available ones. Specifically, uniqueness of the solution to $\textrm{VI}(\Omega(x)\cap\mc Y, F(\cdot, x))$, for fixed $x \in \mc X$, stems from \cite[Ch.~3]{facchinei2003finite}, while the Lipschitz condition is derived from the Dini's theorem \cite{rockafellar2009variational}.   \hfill\qedsymbol

\smallskip

\textit{Proof of Theorem~\ref{th:stability}}:	The feasibility of each iterate in Algorithm~\ref{alg:two_layer} follows immediately by including the state constraints $\mc X$ into $\Omega(\cdot)$, as specified in Remark~\ref{rem:state_constraints}. The convergence of the sequence $\{(x_k, \bs{y}_k(x_k))\}_{k \in \N}$, instead, is a direct consequence of \cite[Th.~4]{lessard2016analysis} after noting that the dissipative inequality in \eqref{eq:IQC} amounts to a pointwise quadratic constraint, parametric in the controller gains $K_i$, $i \in \mc I$, characterizing the feedback interconnection described in Fig.~\ref{fig:control_loop}, for which closed-loop stability can be claimed if $A$ is Schur and \eqref{eq:MI} is verified for some matrix $X \succ 0$ and coefficients $\lambda\geq 0$, $\rho \in [0,1)$. This latter condition on the parameter $\rho$ ensures an exponential convergence rate, as \eqref{eq:MI} implies $\|x_k - x^*\| \le \sqrt{\textrm{cond}(X)} \rho^k \|x_0 - x^*\|$ for all $k \in \N$, where $x^*$ denotes some equilibrium point for the closed-loop system. Specifically, the obtained co-evolutionary equilibrium $((I-A)^{-1} B K \bs{y}^*, \bs{y}^*)$ stems from the standard equilibrium condition with nonlinear controller $\kappa$ in \eqref{eq:controller} with invertible $(I-A)$ as $\Lambda(A) \subset \mc B_1$.	\hfill\qedsymbol
	
\smallskip

\textit{Proof of Proposition~\ref{prop:controller-omega-BMI}}: By imposing $K_i = \omega I_m$ for all $i \in \mc I$, from \eqref{eq:control_opt} we obtain:
\ifTwoColumn
\else
$$
	\left\{
	\begin{aligned}
		&\underset{\omega, X,\lambda}{\textrm{min}} && -\omega\\
		&~\normaltext{\textrm{ s.t. }}&&\begin{bmatrix} A^\top X A-{\rho}^2 X  & \omega A^\top X (B \otimes \bsone_N^\top)   \\ \omega (X (B \otimes \bsone_N^\top))^\top A  & \omega^2 (B \otimes \bsone_N^\top)^\top X (B \otimes \bsone_N^\top) \end{bmatrix} + \lambda \begin{bmatrix}(\bar{\theta}/\eta)^2 I & \phantom{-}\bs 0 \\ \bs 0 & -I \end{bmatrix} \preccurlyeq \bs 0 ,\\
		&&& \omega\in [0,1] , \lambda \ge0, X \in \bbS^n_{\succ 0} ,
	\end{aligned}
	\right.
	$$
\fi
where the constraint $\omega\in [0,1]$ follows directly from $\|K_i\|=\|\omega I_m\| = |\omega| \le 1$ and $[K_i]_{hk} = \omega \ge 0$ for all $i \in \mc I$, while the cost becomes $ \|K-(I_m \otimes \bsone_N^\top)\| = \|(\omega-1)(I_m \otimes \bsone_N^\top)\|= |\omega-1|~\|I_m \otimes \bsone_N^\top\|$ which takes its minimum when $\omega$ approaches its upper bound. The \gls{BMI} reformulation in \eqref{eq:controller-omega-BMI} now follows by defining $\bar B \eqdef B \otimes \bsone_N^\top$, rearranging the matrix inequality above (especially the quadratic terms), and direct application of the Schur's complement.
\hfill\qedsymbol

\smallskip

\textit{Proof of Proposition~\ref{prop:stab}}:	The weighted Laplacian matrix associated with the graph $\mc{G}$, i.e., $L \eqdef D W D^\top$, is known to be symmetric, and so is the scaled matrix $\alpha I_{n} - \tau L$, $\alpha\in(0,1]$: in fact, reverting the sign of the weighted Laplacian matrix, scaling by any $\tau$ and summing it with a scaled identity matrix are all operations that do not alter the symmetry. The symmetry of $A_F$ thus follows by repeating precisely the same reasoning after noting that the weighted Laplacian matrix associated with the subgraph consisting of follower nodes, $L_F \eqdef D_F W D_F^\top$, can also be obtained as $L_F = P_F^\top L P_F$, where $P_F \in \R^{n \times n_F}$ is constructed by eliminating the columns of the scaled identity matrix $\alpha I_n$ that correspond to the influencer nodes.

We now rely on the eigenvalue properties of the sum of Hermitian matrices to claim the result. Specifically, from \cite[Cor.~4.3.15]{horn13} we note that each eigenvalue belonging to the spectrum of the matrix $A_F$, $\Lambda(A_F)$, is bounded as:
$$
	\lambda_i(A_F) \in \left[\underset{i\in\mc M_F}{\textrm{min}} \, \{\alpha_i\} - \tau \lambda_{\textrm{max}}(L_F), \, \underset{i\in\mc M_F}{\textrm{max}} \, \{\alpha_i\} - \tau \lambda_{\textrm{min}}(L_F)\right],
$$
for all $i\in\mc M_F$. Since $\mc{G}$ is connected and $W \succ 0$, from \cite[Lemma 10.36]{mesbahi2010graph} we know that $L_F \succ 0$, and therefore to ensure that $\Lambda(A_F) \subset \mathcal{B}_1$ with $\tau >0$, it suffices to verify $|\lambda_i(A_F)|<1$ for all $i \in \mathcal{M}_F$, that is
$\textrm{min}_{i\in\mc M_F} \, \{\alpha_i\} - \tau \lambda_{\textrm{max}}(L_F)>-1$ and $\textrm{max}_{i\in\mc M_F} \, \{\alpha_i\}-\tau \lambda_{\textrm{min}}(L_F)< 1$.
While this latter relation is directly implied by the conditions $\alpha_i \leq 1$, $\tau > 0$ and $L_F \succ 0$, the former requires one to impose $\tau \in (0, \textrm{min}_{i \in \mathcal{M}_F}\{1+\alpha_i\}/\lambda_{\textrm{max}}(L_F))$, from which the thesis holds true. 
\hfill\qedsymbol

\smallskip

\textit{Proof of Theorem~\ref{th:CL_S_proc}}: Consider any co-evolutionary equilibrium of the \normaltext{\gls{GNEP}} $\Gamma$ in \eqref{eq:single_prob_example_schematic} and \normaltext{\gls{LTI}} system in \eqref{eq:LTI_network_aug}, $(x_F^*,\bs y^*)$. This latter reflects onto the augmented dynamics \eqref{eq:LTI_Delta} as:
$$
	\left\{
	\begin{aligned}
		& x^*_F = \mathcal{A} x^*_F + \mathcal{B}\gamma^*,\\
		&\zeta^* = \mathcal{C} x^*_F + \mathcal{D} u^* =  \mathcal{C} x^*_F + \mathcal{D}  \bs y(x^*_F),\\
		&\gamma^* = \Delta \zeta^*,
	\end{aligned}
	\right.
$$
where we have implicitly recalled that $u=\bs y(x^*_F)$. Let us then consider the expression in~\eqref{end.lmi}. After pre- and post-multiplying that matrix inequality with vector $\textrm{col}(e_F, \gamma-\gamma^*, u-u^*)$, where $e_F \eqdef x_F-x_F^*$, using the first relation above, we directly obtain:
\ifTwoColumn
\else
$$
	\!\!\!(e^+_F)^\top X e^+_F \leq \rho^2 (e_F)^\top X e_F + \lambda(\|u-u^*\|^2- (\hat\theta/\eta)^2 \|e_F\|^2)  - (\star)^\top \begin{bmatrix} R & S \\ S^\top & T  
	\end{bmatrix} \begin{bmatrix} \zeta-\zeta^* \\
		\gamma-\gamma^*
	\end{bmatrix},
$$
\fi
where $e_F^+ = \mathcal{A} x_F + \mc B \gamma - x_F^*$. Thus, in view of the quadratic constraint~\eqref{eq:IQC} and the fact that $\lambda\ge0$, the term $\lambda(\|u-u^*\|^2- (\hat\theta/\eta)^2 \|e_F\|^2)$ is non-positive and therefore it can be neglected. For the last term, by substituting $\gamma = \Delta \zeta$ from \eqref{eq:LTI_Delta}, we obtain:
\ifTwoColumn
\else
	$$
		-(\star)^\top \begin{bmatrix} R & S \\ S^\top & T  
	\end{bmatrix} \begin{bmatrix} \zeta-\zeta^* \\
		\gamma-\gamma^*
	\end{bmatrix} = -(\star)^\top \begin{bmatrix} R & S \\ S^\top & T 
	\end{bmatrix} \left(\begin{bmatrix} I_{2n_F}\\ \Delta \end{bmatrix} \begin{bmatrix} \zeta-\zeta^*\end{bmatrix}\right)
	$$
\fi
which is required to be negative by~\eqref{start.lmi}, and hence it can be neglected as well, yielding the contraction $(e^+_F)^\top X e^+_F \leq \rho^2 (e_F)^\top X e_F$ since $\rho \in [0,1)$. This ensures closed-loop stability, and specifically we have:
$$
\|x_{F,k}-x_F^*\| \leq \sqrt{\textrm{cond}(X)} \rho^k \|x_{F,0}-x_F^*\|, 
$$
i.e., the \normaltext{\gls{GNEP}} \eqref{eq:single_prob_example_schematic} and dynamics \eqref{eq:LTI_network_aug} co-evolve to some equilibrium  $((I-A_F)^{-1} \bar B_F \bs{y}^*, \bs{y}^*)$ exponentially fast, where $(I-A_F)$ is invertible since $\tau \in (0, (1+\alpha)/\lambda_\normaltext{\textrm{max}}(L_F))$ guarantees that $\Lambda(A_F) \subset \mc B_1$ in view of Proposition~\ref{prop:stab}. The feasibility of each iterate in Algorithm~\ref{alg:two_layer} follows by adopting the same arguments as in the proof of Theorem~\ref{th:stability}.
\hfill\qedsymbol

\smallskip

\textit{Proof of Theorem~\ref{th:decomposition}}:
The derivation of the condition in \eqref{end.lmi2} follows directly from the properties of the Kronecker product once plugged in~\eqref{end.lmi} the expressions for the decision variables in the statement of the theorem. In particular, by representing~\eqref{end.lmi} as $(\star)^\top (M_1 \otimes I_{n_F}) (M_2 \otimes I_{n_F}) + \lambda M_3 \otimes I_{n_F} \preccurlyeq 0 $, for appropriate block-matrices $M_1$, $M_2$ and $M_3$, we obtain:
\ifTwoColumn
\else
\begin{align}
	&(\star)^\top \!\!\left[\begin{array}{c|c} \begin{array}{cc} {\chi} & \phantom{-}0 \\ 0 & -{\chi}\end{array} & {\bf 0}\\ \hline {\bf 0} & \begin{array}{ccc} {r_1} & \bs 0 & \bs 0 \\ \bs 0 & r_2 I_N & \bs 0\\ \bs 0 & \bs 0 & {t}\end{array}  
	\end{array}\right] \!\!\left[\begin{array}{ccc} 1  &\!\! \bsone_2^\top &\!\! \bs 0   \\ 
		{\rho}  &\!\!  \bs 0 &\!\! \bs 0\\ \hline
		\begin{array}{c} 1 \\ \bs 0 \end{array} &\!\! \begin{array}{c} \bs 0 \\ \bs 0 \end{array} &\!\! \begin{array}{c} \bs 0 \\ I_N \end{array} \\
		\bs 0 &\!\! I_{2} &\!\! \bs 0
	\end{array}\right]+{\lambda} \begin{bmatrix} (\hat\theta/\eta)^2 & \bs 0 & \phantom{-}\bs 0\\ \bs 0 & \bs 0 & \phantom{-}\bs 0 \\ \bs 0 & \bs 0& -I_N  
	\end{bmatrix} \preccurlyeq \bs0. \label{end.lmi.dummy}
\end{align}  
\fi
Now, developing the lowest diagonal block in~\eqref{end.lmi.dummy}, we obtain,
\ifTwoColumn
\else
\begin{align*}
	&(\star)^\top \!\!\left[\begin{array}{cc}  r_2 I_N & \bs 0\\ \bs 0 & {t}\end{array} \right] \!\!\left[\begin{array}{cc}  \bs 0 & I_N \\ I_2 & \bs 0  \end{array}\right] + {\lambda} \begin{bmatrix}  \bs 0 & \bs 0 \\ \bs 0 &  -I_N  
	\end{bmatrix} \preccurlyeq \bs 0 \iff \left[\begin{array}{cc}  t & \bs 0\\ \bs 0 & ({r}_2 -\lambda)I_N\end{array} \right] \preccurlyeq \bs 0 \,\\
	&\iff \left[\begin{array}{cc}  t & \bs 0\\ \bs 0 & {r}_2 -\lambda\end{array} \right] \preccurlyeq \bs 0 \iff(\star)^\top \!\!\left[\begin{array}{cc}  r_2 & \bs 0\\ \bs 0 & {t}\end{array} \right] \!\!\left[\begin{array}{cc}  \bs 0 & 1 \\ I_2 & \bs 0  \end{array}\right] + {\lambda} \begin{bmatrix}  \bs 0 & \bs 0 \\ \bs 0 &  -1  
	\end{bmatrix} \preccurlyeq \bs 0,
\end{align*}  
\fi
from which condition~\eqref{end.lmi2} follows.
For what concern instead the condition in \eqref{start.lmi2}, we rewrite~\eqref{start.lmi} as
\begin{eqnarray}
	&& r_1 I_{n_F} + t_1 \tau^2 L_F^\top L_F \succ \bs 0  \label{eq.dd} \\
	&& r_2 I_{n_F N} + t_2 \bar B_F^\top \bar B_F \succ \bs 0.\nonumber
\end{eqnarray}
Following the procedure described in \cite[Th.~5]{Massioni2014}, we perform a singular value decomposition for both $\tau L_F = U_1^\top \Sigma_1 V_1$ and $\bar B_F = U_2^\top \Sigma_2 V_2$, which yields the following relations (for $\tau L_F$, though identical calculations can be performed with $\bar B_F$):
\begin{align}
	\eqref{eq.dd} & \iff r_1 I_{n_F} + t_1 V_1^\top \Sigma_1^2V_1 \succ \bs0, \nonumber\\
	&~~\cong ~~~ r_1 I_{n_F} + t_1  \Sigma_1^2 \succ \bs0, \nonumber\\
	& \iff r_1 + t_1 \sigma_i^2(\tau L_F) > 0,~\forall i \in \{1, \ldots, n_F\} \label{eq.dd2},
\end{align}
where, in this case, $\sigma_i(\cdot)$ denotes the $i$-th singular value of its argument. Then, since each $\sigma_i^2(\tau L_F) \in [0, \delta_{\normaltext{\textrm{max}},1}]$, we obtain:
\begin{equation*}
	\eqref{eq.dd2} \iff r_1>0,~r_1 + t_1 \delta_{\normaltext{\textrm{max}},1}^2 > 0.
\end{equation*}
The claim follows after applying the same procedure to $\bar B_F$. 
\hfill\qedsymbol

\bibliographystyle{IEEEtran}
\bibliography{GNEP_LTIsys.bib}



\end{document}